\documentclass[aps,prb,floatfix,showpacs,superscriptaddress, twocolumn, footinbib]{revtex4-2}

\usepackage{amsmath}
\usepackage{graphicx}
\usepackage{dcolumn}
\newcolumntype{d}[1]{D{.}{.}{#1}}
\usepackage{bm}
\usepackage{epstopdf}
\usepackage{stmaryrd}
\usepackage{tikz}
\usepackage{lipsum}
\usepackage{pgf}
\usepackage{float}
\usepackage{hyperref}
\usepackage{tabularx}
\usepackage{appendix}
\usepackage[section]{placeins}
\usepackage[none]{hyphenat}

\usepackage{subfigure}

\definecolor{lightblue}{RGB}{0,170,255}

\newcommand{\be}{\begin{equation}} 
\newcommand{\ee}{\end{equation}} 
\newcommand{\bea}{\begin{eqnarray}} 
\newcommand{\eea}{\end{eqnarray}} 
\newcommand{\bqa}{\begin{eqnarray}}
\newcommand{\eqa}{\end{eqnarray}}

\newcommand* {\bra}[1]{\langle {#1} |}
\newcommand* {\ket}[1]{| {#1} \rangle}

\newcommand{\appropto}{\mathrel{\vcenter{
			\offinterlineskip\halign{\hfil$##$\cr
				\propto\cr\noalign{\kern2pt}\sim\cr\noalign{\kern-2pt}}}}}

\usepackage[normalem]{ulem}

\newcommand{\ba}{\arraycolsep 0.3ex \begin{array}{rl}}
	\newcommand{\ea}{\end{array}}
\newcommand{\bc}{\begin{cases}}
	\newcommand{\ec}{\end{cases}}
\usepackage{color}
\newcolumntype{C}[1]{>{\centering\arraybackslash}p{#1}}

\begin{document}

\title {Topological defect-mediated corner states and higher-order bulk topology in a 2D crystalline insulator }

\author{Manideep Gone}
\email{22pph001@lnmiit.ac.in}
\affiliation{Department of Physics, The LNM-Institute of Information Technology, Jaipur 302031, India}

\author{Srijata Lahiri}
\email{srijata.lahiri@iitg.ac.in}
\affiliation{Department of Physics, Indian Institute of Technology Guwahati, Guwahati 781039, India}

\author{Nabyendu Das}
\email{nabyendu@lnmiit.ac.in}
\affiliation{Department of Physics, The LNM-Institute of Information Technology, Jaipur 302031, India}

\date{\today}

\begin{abstract}
We report the appearance of non-trivial zero-energy corner modes in the form of topological defects (trimers) in a carefully designed 2D crystalline topological insulator. The proposed scenario is developed via an unconventional stacking of 1D topological atomic chains with crystalline mirror symmetry along the diagonal $(y=x)$ line. Our analysis shows that by systematically varying the hopping parameters $t$ (intra-chain), $v$ (within the unit cell) and $w$ (between the unit cells), the system exhibits more than one distinct non-trivial second-order topological phases. These phases are distinguished by the zero-energy corner modes. In one of these phases, the system supports four zero modes. Two of them reside on the trimers, and the rest on isolated sites situated at the corners along the diagonal line. However, in the second case, the zero modes on the isolated sites persist at the corners while the zero modes on the trimers vanish. A critical look at the phase evolution of the Bloch states helps in investigating the topology of these phases by using winding numbers. Our work also shows the bulk-corner correspondence that exists between the invariants and the zero modes at the corners. With four zero modes at the corners and  a winding number as $2$, we conclude that the system has transformed into  a second-order topological insulator via tuning of the hopping amplitudes. 
\end{abstract}

\maketitle

\label{sec:introducion}
\section{Introduction}   

Topological phases (TPs) \cite{1.1.Kosterlitz1972,1.2.Kosterlitz1973,2.1.PhysRevLett.50.1153,2.2.PhysRevLett.49.405} do not rely on symmetry-breaking concepts or a local parameter to define phase transitions, unlike the Landau paradigm \cite{LandauLifshitz1980}. These phases are defined by the topology of electronic band structures and remain protected by symmetries of the system. The novel manifestation of the intricate relation between the symmetries in the system and the topology of the bulk is seen through bulk-boundary correspondence in these systems. Based on the symmetries and topology, Altland-Zirnbauer proposed a tenfold way to classify various topological materials \cite{PhysRevB.55.1142}. These classifications are experimentally well established through integer quantum Hall systems \cite{3.PhysRevLett.45.494,4.PhysRevLett.49.405}, fractional quantum Hall systems \cite{5.RevModPhys.71.S298}, topological insulators (TIs) \cite{6.PhysRevLett.89.077002,7.Moore_2010,8.Ando_2013}, and topological superconductors \cite{9.Schnyder_2008,10.Kitaev_2009,11.Read_2000,12.Kitaev_2001,13.Sato_2017}. TIs are commonly recognized for having conducting boundaries with insulating bulk, and bulk topological invariants are used to distinguish the TPs. The Qi-Wu-Zhang (QWZ) model in 2D, which has Chiral boundary states and Chern number as a bulk topological invariant to distinguish the TPs \cite{14.Asb_th_2016,15.Qi_2006}. The Bernevig-Hughes-Zhang (BHZ) model \cite{16.Bernevig_2006}, a 2D TI with band-inversion supports helical edge states protected by time-reversal symmetry, gives rise to the quantum spin Hall effect.  The Kane-Mele model \cite{17.PhysRevLett.95.226801,18.PhysRevLett.95.146802} extends the graphene model by including intrinsic spin-orbit coupling, which also results in quantum spin Hall effects with helical edges. Apart from these materials, there are other topological materials where TPs are classified by zero-energy modes at boundaries rather than conducting ones. The Su-Schrieffer-Heeger (SSH) model in 1D is the simplest of all, which describes spinless particles in the bipartite lattice having zero energy modes at the edges and winding number as a bulk topological invariant, which distinguishes the TPs present in the system \cite{19.PhysRevLett.42.1698,20.PhysRevB.21.2388,21.PhysRevB.22.2099,22.RevModPhys.60.781,23.Quantal_phases,24.PhysRevLett.62.2747}.

In recent years, there have been ongoing efforts to generalize the concept of TPs from lower-order to higher-order states, where the topological characteristics observed in lower dimensions are considered mathematical as well as physical generalizations. As an illustration, dipole polarization represents a measure that can be defined using the quantized topological property found in one-dimensional crystalline insulators, extending to multi-pole polarization with underlying topological roots \cite{22.RevModPhys.60.781}. The system with these novel higher-order TPs, especially the higher-order topological insulators (HOTIs), exhibits gapless states in their co-dimension-$1$. The edge or surface spectrum becomes gapped \cite{25.Benalcazar_2017,26.Khalaf_2018,27.PhysRevLett.119.246401,28.PhysRevLett.119.246402,29.Schindler2018,30.Dutt_2020}. 
In higher-order topological superconductors, the presence of hinge states in the form of the Majorana modes \cite{31.PhysRevB.104.134508,32.PhysRevLett.121.196801,33.PhysRevResearch.2.032068,34.PhysRevB.105.195106} allows for topological quantum computing \cite{35.PhysRevResearch.2.043025,36.RevModPhys.80.1083}.\\

Recent studies also support the emergence of higher-order TPs from stacking lower-dimensional topological atomic  chains, where higher-order TPs are observed \cite{37.Zhu2020,38.PhysRevB.107.045118,PhysRevB.98.205147,SerraGarcia2018,PhysRevB.101.241104,PhysRevLett.108.220401,PhysRevLett.109.116404}.
Numerous studies have examined the various 1D topological atomic  model extensions into two dimensions by considering more than two sub-lattices per unit cell, which results in various interesting TPs protected by the symmetries of the topological atomic  chain \cite{39.PhysRevResearch.4.023193,40.PhysRevB.100.075437,41.PhysRevB.106.245109}. There are other 2D extensions possible by coupling the  topological atomic  chains to each other so that an effective 2D model is created by stacking the  topological atomic  chains with different intra-chain and inter-chain hopping amplitudes by protecting the symmetries of the individual stacked topological atomic  chains \cite{42.Het_nyi_2018,43.Master_thesis,44.Rosenberg_2022,45.PhysRevB.104.134511,46.PhysRevB.100.104522} or by breaking any of the symmetries \cite{47.Agrawal_2023}. Depending on the nature of extensions, distinct TPs arise in the system and to classify these TPs and establish a bulk-boundary correspondence, different topological invariants are used. For most of the 2D system, the Chern number is used as a topological invariant \cite{48.PhysRevLett.61.2015,49.3Sticlet_2012,50.PhysRevB.106.165423,51.Tarnowski_2019,52.doi:10.1143/JPSJ.74.1674,53.1984}. But when the Berry curvature is zero \cite{63.Liu_2017,64.PhysRevB.97.035442} other topological invariants, like winding number \cite{54.Malakar_2023,55.pérezgonzález2018sshmodellongrangehoppings,56.Ghosh_2023,57.PhysRevB.103.075117,58.PhysRevB.104.075113}, mirror-graded topological invariant \cite{59.Kariyado2017,60.Lau_2016,61.Zhang2013}, and diagonal winding number  \cite{62.Seshadri_2019}  are used to establish a connection with the boundary modes.  The quantum spin Hall systems in 3D offer a framework to study 2D layered systems, resulting in different kinds of TPs. These phases are classified as strong TIs, weak TIs and Topological semimetals \cite{65.PhysRevLett.98.106803,66.PhysRevB.75.121306,67.PhysRevLett.95.226801,68.PhysRevLett.95.146802,69.PhysRevB.86.045102,70.PhysRevLett.116.066801,71.Burkov2016,72.Bernevig2018}. We were initially motivated by the study of Chen, Bo-Hung on a two-dimensional extended topological atomic  model \cite{43.Master_thesis} and proposed a second-order topological insulator (SOTI) with exotic second-order topological phases (SOTPs). These stacking are done in the y-direction by an inter-chain hopping or a staggered hopping with $v$ and $w$, or stacking is done by shifting of chains by a unit cell in each layer in the y-direction. Both these kinds of extensions preserve the existing symmetries of the topological atomic  chain.\\

In this work, we extensively studied a non-trivial stacking mechanism which introduces additional crystalline mirror symmetry and also symmetry-protected topological corner defects along the $y=x$ line. These topological corner defects are absent in conventional stacking mechanisms \cite{37.Zhu2020,38.PhysRevB.107.045118,PhysRevB.98.205147,SerraGarcia2018,PhysRevB.101.241104,39.PhysRevResearch.4.023193,40.PhysRevB.100.075437,41.PhysRevB.106.245109,42.Het_nyi_2018,44.Rosenberg_2022,45.PhysRevB.104.134511,46.PhysRevB.100.104522,47.Agrawal_2023} because their stacking arrangement and boundary termination suppresses the formation of these defects. Our stacking mechanism intrinsically terminates the boundary, hosting these topological defects along the corners of $y=x$ line leading to interesting bulk-corner correspondence.
In section \ref{sec: The 2D extension of the topological atomic  model}, we briefly explain the non-trivial nature of the stacking, followed by the Hamiltonian with  energy spectrum to identify the signatures of TIs. A bulk invariant to classify the higher-order phases of the model, as well as bulk polarization and symmetries of the system. By finite boundary conditions, possible kinds of effective edges are highlighted in Section  \ref{sec: The edge of the system}. Section \ref{sec: Bulk-Boundary Correspondace} argues how these finite boundary conditions failed to capture the significance of the local defects at the corners and finally ends by establishing bulk-corner correspondence with SOTPs. In the last section, we conclude with our remarks on symmetry aspects of the bulk-boundary correspondence in this system and future directions for further studies.
 
 \section {The 2D extension of the topological atomic  model}
\label{sec: The 2D extension of the topological atomic  model}

\begin{figure}[htpb]
     \centering
     \includegraphics[width=0.5\textwidth]{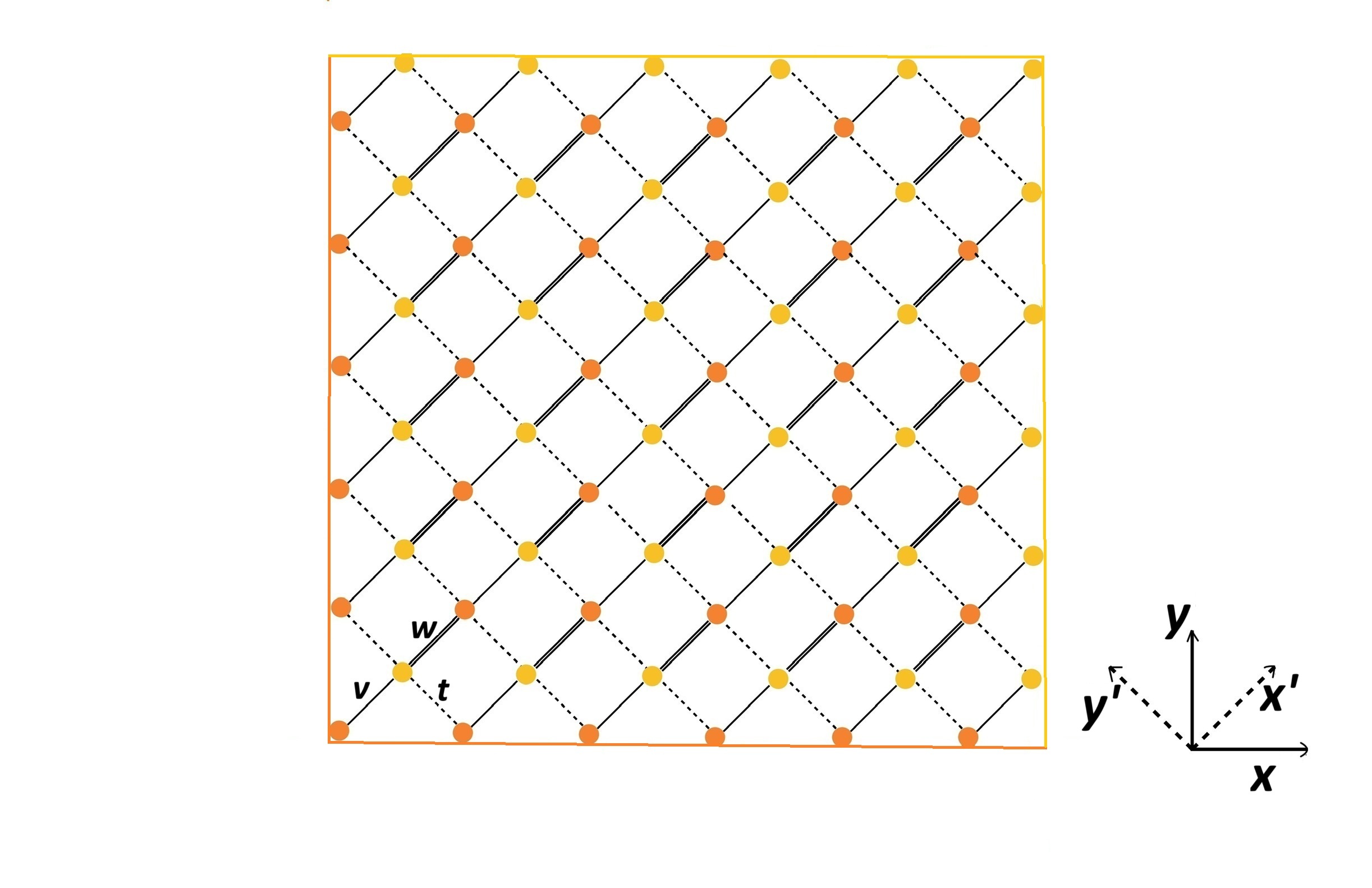}
     \caption{Illustration of the 2D crystalline model formed by stacking  1D topological atomic chains by orienting them along the line $y=x$, resulting in two different pairs of edges formed from A and B atoms. Orange and yellow dots represent the two different atoms, A and B. The thin lines represent the hopping amplitude $v$(within the unit cell). The solid and dashed lines represent the NN  hopping amplitude  $w$(within the chain) and  $t$(between the chains), respectively. Orange and yellow lines indicated the different edges of the model.}
     \label{fig:fig1}
 \end{figure}

 This section investigates a two-dimensional crystalline system formed by stacking topological atomic chains. These chains are formed by atoms with spinless particles arranged in one-dimensional lattice, where the electronic properties are governed by the global topological properties of the system. These topological atomic chains are oriented along the $y=x$ diagonal direction, as illustrated in Figure \ref{fig:fig1}. When layered, these chains are shifted and reduced by a unit cell in relative to the previous chain in the stacking direction. Thus the resulting configuration appears as a two-dimensional periodic square lattice with two sites A (orange) and B (yellow) per unit cell. We consider spinless particles with three different hopping parameters: $v$ for intra-cell hopping (thin line), $w$ for the nearest-neighbor (NN) inter-cell hopping in the same chain (solid line)  and  $t$ for NN inter-chain hopping (dashed lines). Hopping from A to A  or B to B sites is not allowed in the system. This stacking can also be considered as topological atomic chains organized in the x-direction, with intra-cell hopping $v$ and inter-cell hopping $t$ with $N_{x}$ lattice points. Such $N_{y}$ chains are layered in the y-direction, with alternate inter-chain hopping strengths as $t$ and $w$. Alternatively this can be considered as analogous stacking along x-direction when the chains are oriented in y-directions.\\
 
 The non-trivial nature of stacking is due to the orientation of chains and the shifting and deliberate reduction of one unit cell relative to the prior chain in the stacking direction. This kind of orientation introduces crystalline mirror symmetry along the $y=x$ line. Also, it features the significance of the combined effects of $k_x$ and $k_y$ in defining the bulk topological properties when the periodic conditions are imposed along the x and y-directions. The unit-cell shift preserves the topological chains' chiral symmetry. It also introduces inter-chain hopping $t$ as NN rather than the next-nearest-neighbor (NNN) hopping term, ensuring no long-range hopping terms in the system. Apart from shifting, each chain is reduced by one unit cell relative to the preceding layer. This design creates an intrinsic boundary termination, which leads to the formation of topological defects along the corners of the $y=x$ line. These defects remain protected by the system's crystalline mirror symmetry and chiral symmetry. The real-space Hamiltonian for this model is as follows:

 \begin{equation}\label{eq:1}
        \centering
        \begin{aligned}
          \hat{H}& = \sum_{m_x}^{N_x}\sum_{m_y}^{N_y} v\ket{m_x,m_y,A}\bra{m_x,m_y,B}\\ &+\sum_{m_x}^{N_x-1}\sum_{m_y}^{N_y-1} w\ket{m_x+1,m_y+1,A}\bra{m_x,m_y,B} \\
          &+\sum_{m_x}^{N_x-1}\sum_{m_y}^{N_y} t\ket{m_x+1,m_y,A}\bra{m_x,m_y,B} \\
           &+\sum_{m_x}^{N_x}\sum_{m_y}^{N_y-1} t\ket{m_x,m_y+1,A}\bra{m_x,m_y,B} +h.c.,
        \end{aligned}
\end{equation}
where $m_{x}$ and $m_{y}$ represent the coordinates of the unit cells in x and y directions, respectively; $N_{x}$ and $N_{y}$ represent the number of unit cells in x and y directions, respectively.

\subsection{Bulk Hamiltonian}
In the thermodynamic limit where $N\xrightarrow{}\infty$, the impact of the boundary is negligible, and the physical properties are purely governed by the bulk. We impose complete periodic boundary conditions (CPBC) via Born-von-Karman boundary conditions in both spatial directions by embedding  the system on a 2D toroidal surface. The system results in translation invariance, where the real space Hamiltonian is Fourier transformed to crystal momentum. The Brillouin zone (BZ) (parametric space) is a 2D toroidal surface with the crystal momentum $\mathbf{k}=(k_x,k_y)$ ranging from $[-\pi,\pi]$ in both directions. The bulk Hamiltonian in momentum space  basis $\ket{k_x,k_y,A}$ and $\ket{k_x,k_y,B}$ is given as,

 \begin{equation}\label{eq:2}
        H(\mathbf{k})=\begin{pmatrix}
 0 & h^{*}(k_x,k_y)\\
               h(k_x,k_y) & 0
              \end{pmatrix},
\end{equation}
where $h(k_x,k_y)=v+te^{ik_x}+te^{ik_y}+we^{i(k_x+k_y)}$. Diagonalizing the Hamiltonian $H(\mathbf{k})$ in equation \ref{eq:2} gives energy dispersion and the corresponding Bloch states (energy eigenfunctions) of the bulk Hamiltonian. The energy dispersion is given by 
\begin{equation}\label{eq:3}
    \begin{aligned}
        E_{\pm}(k_x,k_y)&=\pm\sqrt{v^2+w^2+2t^2+f(k_x,k_y)},\\
        f(k_x,k_y)&=2t(v+w)(\cos(k_x)+\cos(k_y))\\
        &+2t^2\cos(k_x-k_y)+2vw\cos(k_x+k_y).
    \end{aligned}
\end{equation}
The Bloch states are given as
\begin{equation}\label{eq:4}
      \ket{\pm u_{\mathbf{k}}}=\frac{1}{\sqrt{2}}\begin{pmatrix}
            \pm e^{-i\phi(k_x,k_y)}\\
            1
        \end{pmatrix},
\end{equation}
where,
\begin{equation}\label{eq:5}
    \begin{aligned}
      \phi(k_x,k_y)&= \tan^{-1}\left( \frac{h_y(k_x,k_y)}{h_x(k_x,k_y)}\right ),  
    \end{aligned}
\end{equation}

\begin{equation}\label{eq:6}
    \begin{aligned}
      h_x(k_x,k_y)&=v+t\cos(k_x)+t\cos(k_y)+w\cos(k_x+k_y),\\
      h_y(k_x,k_y)&=t\sin(k_x)+t\sin(k_y)+w\sin(k_x+k_y).\\
    \end{aligned}
\end{equation}

\begin{figure}[htpb] 

    \begin{minipage}{0.5\textwidth}
        \includegraphics[width=0.45\linewidth,height=0.4\linewidth]{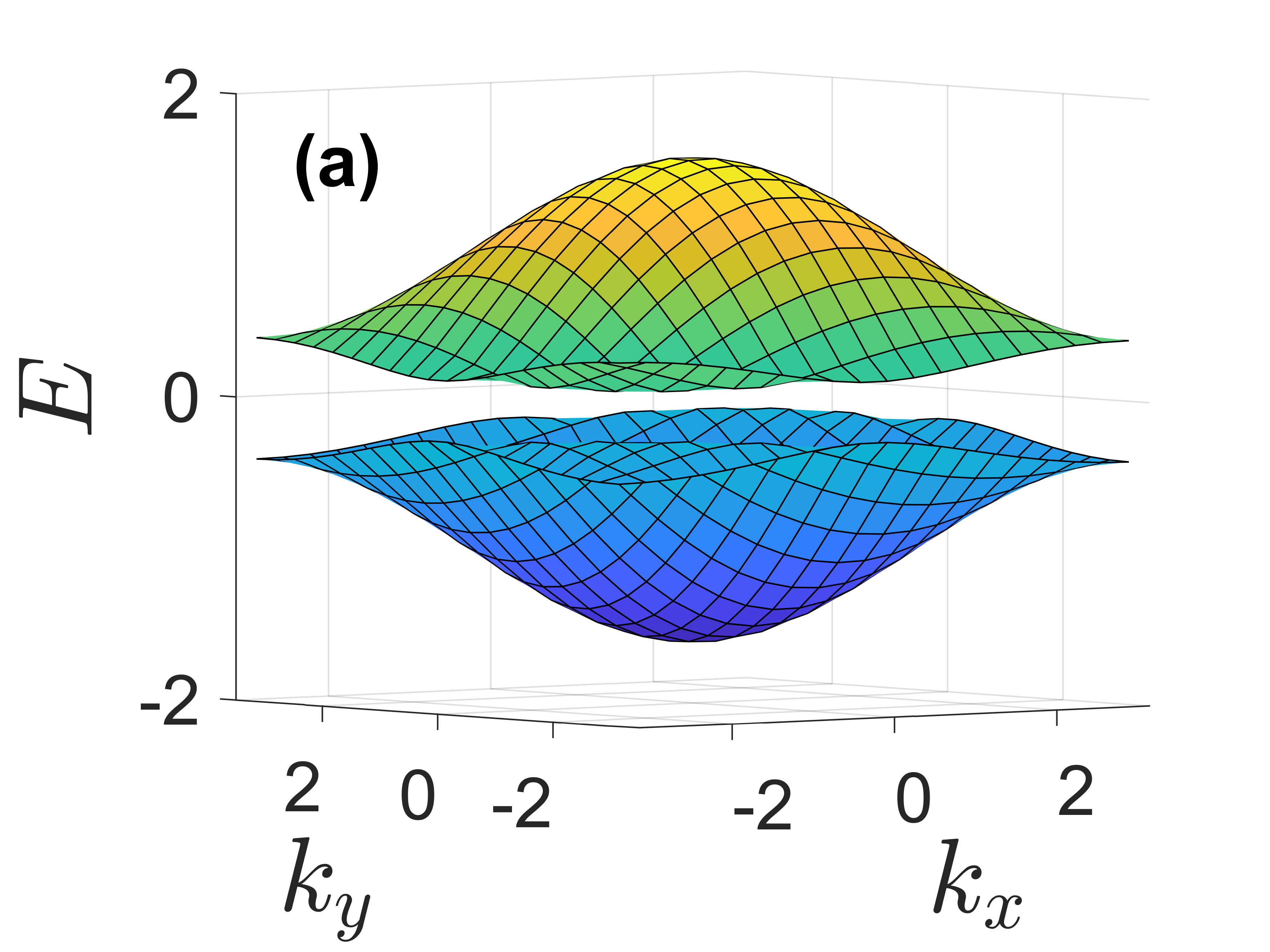}
        \hspace{10pt}
        \includegraphics[width=0.45\linewidth,height=0.4\linewidth]{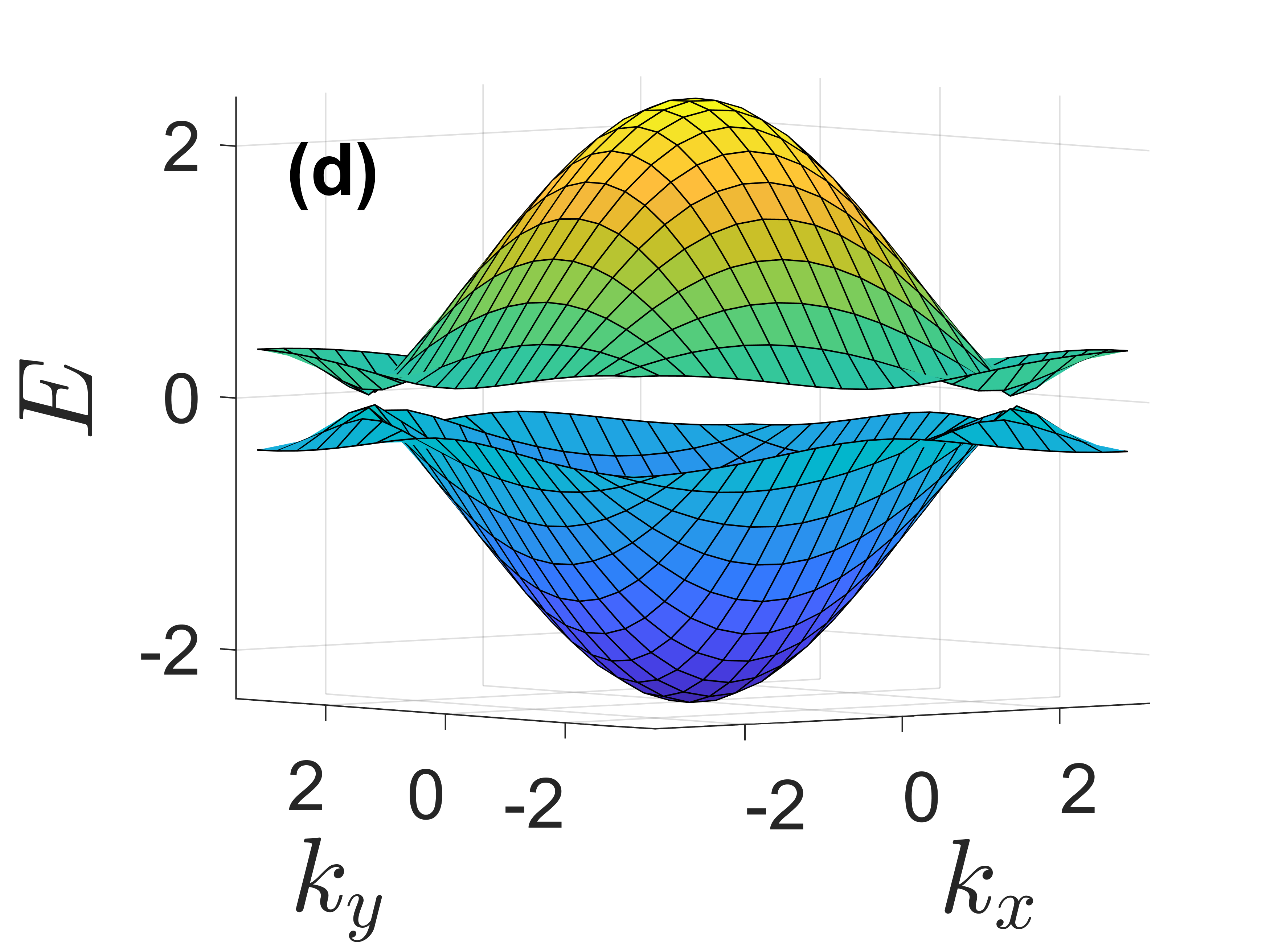}
\end{minipage}

\begin{minipage}{0.5\textwidth}
    \includegraphics[width=0.45\linewidth,height=0.4\linewidth]{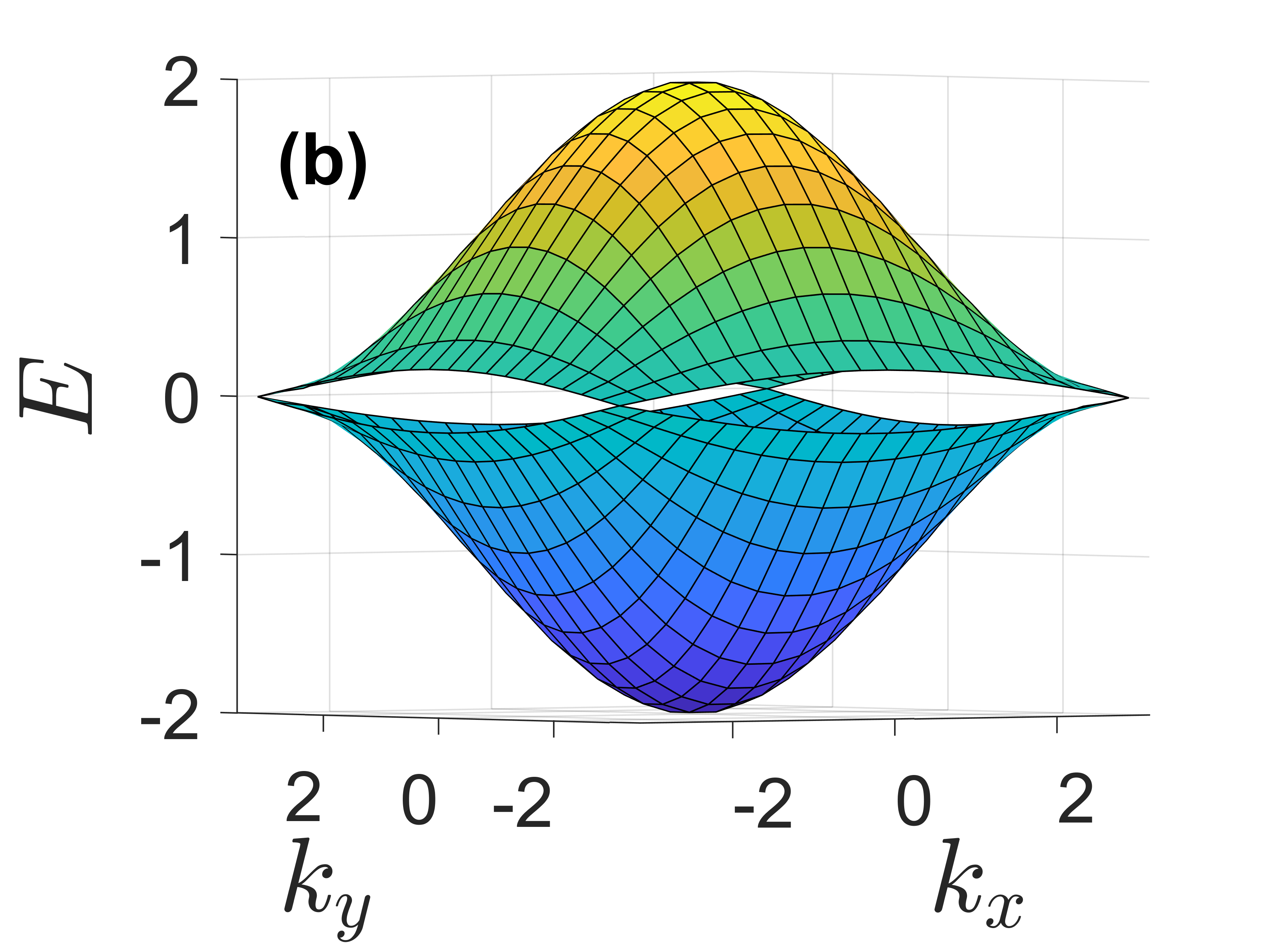}
        \hspace{10pt}
        \includegraphics[width=0.45\linewidth,height=0.4\linewidth]{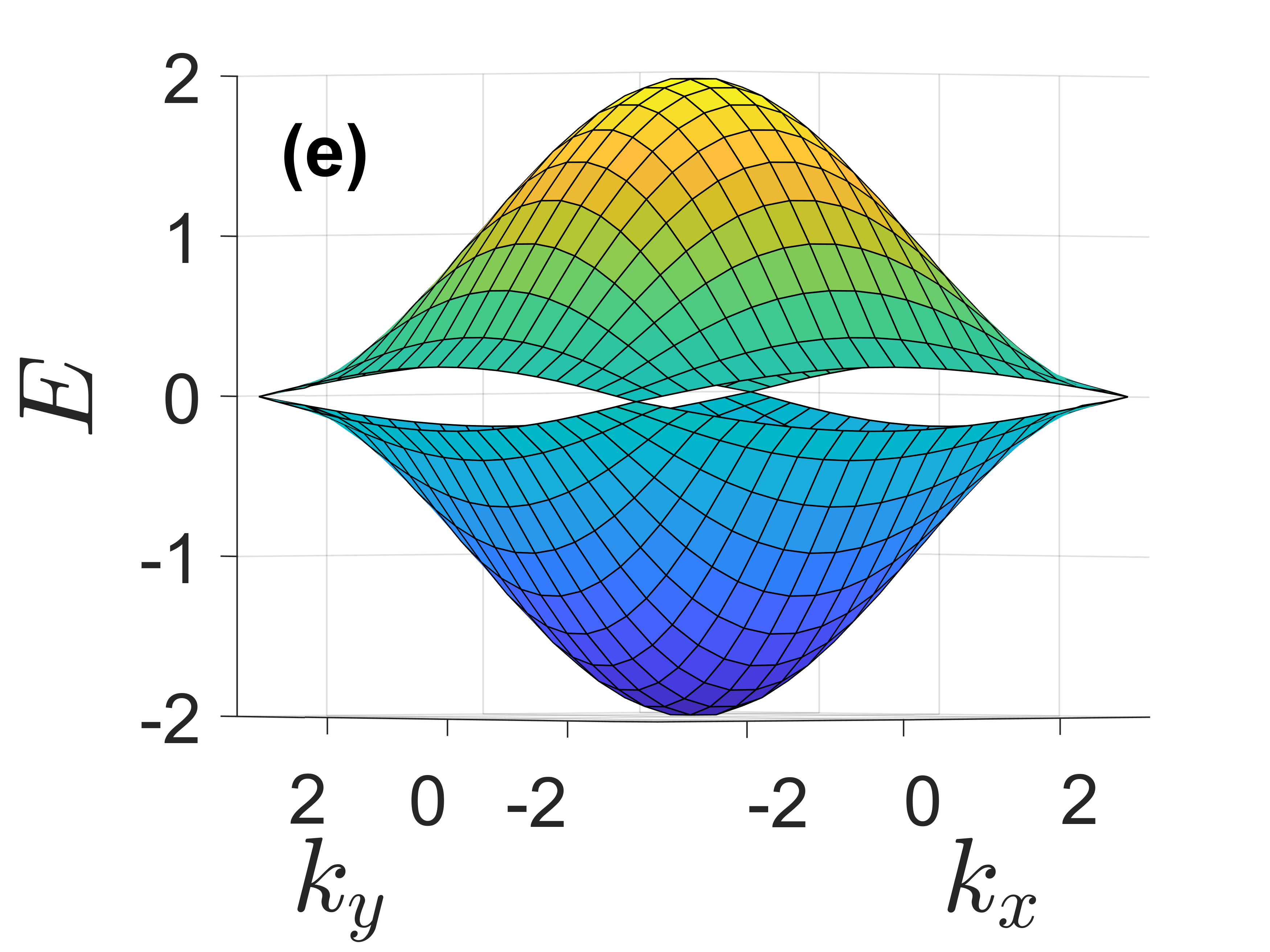}
\end{minipage}

\begin{minipage}{0.5\textwidth}
    \includegraphics[width=0.45\linewidth,height=0.4\linewidth]{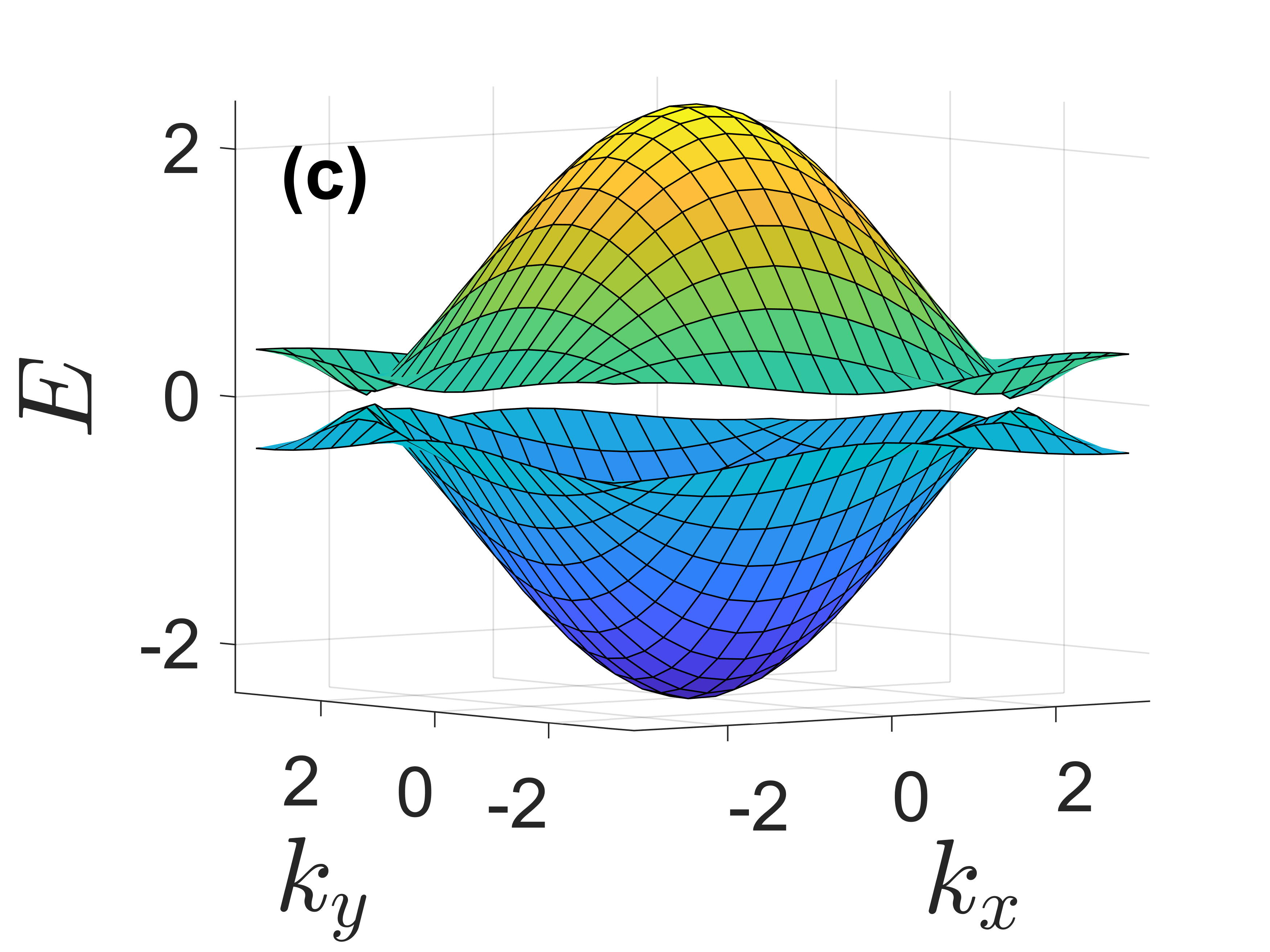}
    \hspace{10pt}
    \includegraphics[width=0.45\linewidth,height=0.4\linewidth]{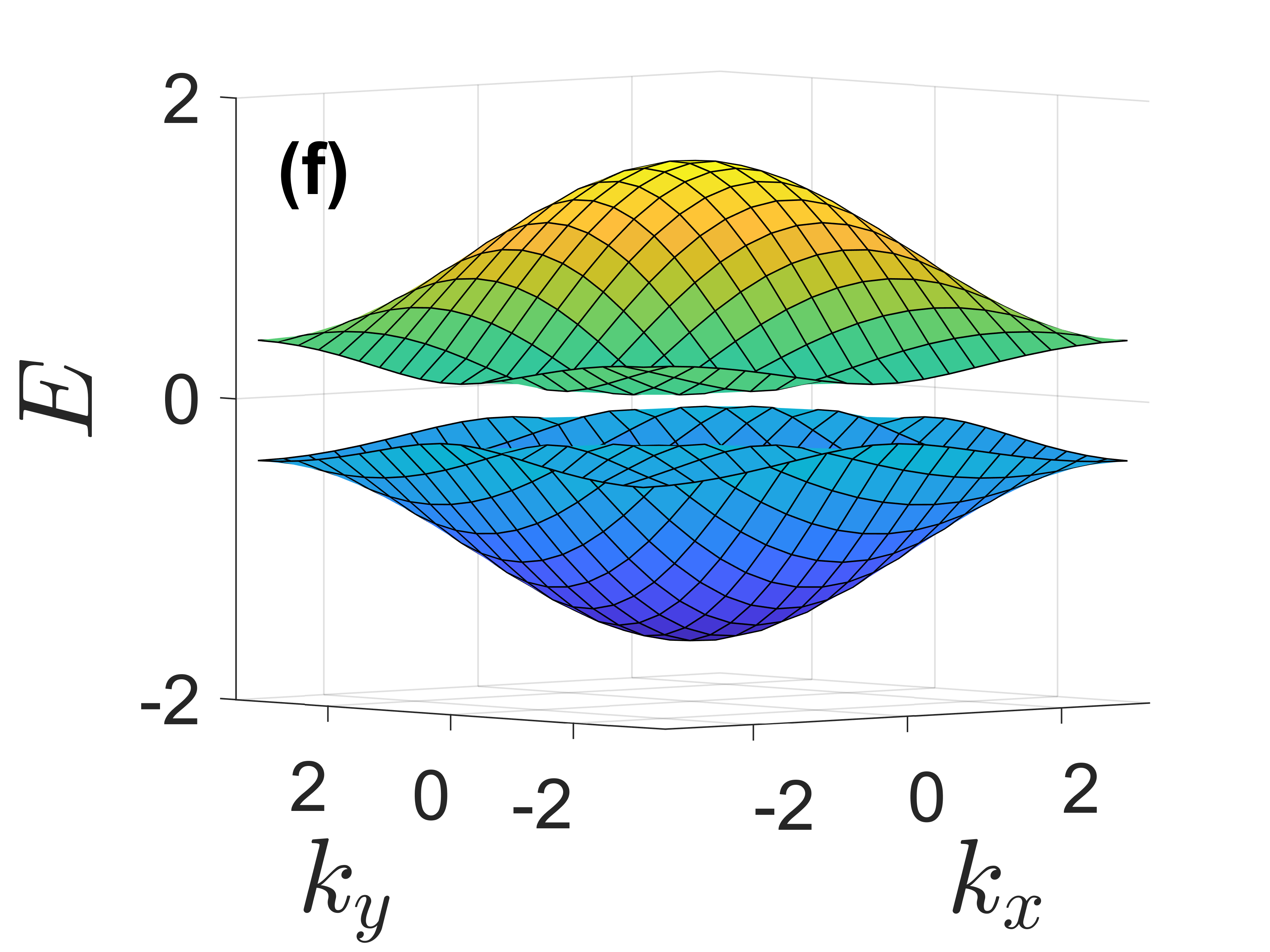}
\end{minipage}

\caption{Graphical representation of electronic band structures in the parametric space $(k_x,k_y)$ under CPBC. Plots (a) to (c) represent the case when the individual topological atomic chains are in a non-trivial topological phase $(v<w)$ with (a) $t<\frac{v+w}{2}$, (b)  $t=\frac{v+w}{2}$ (gap closes) and (c) $t>\frac{v+w}{2}$. Plots (d) to (f) represent the case when the individual topological atomic chains are in a trivial topological phase $(v>w)$ with (d) $t>\frac{v+w}{2}$, (e)  $t=\frac{v+w}{2}$ (gap closes) and (f) $t<\frac{v+w}{2}$.}
\label{fig:2}
\end{figure}

The dispersion relation corresponding to six different choices of intra and inter-cell hopping $(v,w)$, and inter-chain hopping $t$ are shown in Figure \ref{fig:2}. These configurations are $v<w$ and $v>w$ with three regimes of inter-chain hopping: $t<\frac{v+w}{2}$, $t=\frac{v+w}{2}$ and $t>\frac{v+w}{2}$. In the cases $v<w$ and $v>w$, a gap closing occurs when $t=\frac{v+w}{2}$ at four different points in the BZ which are $\delta_1=(-\pi,-\pi)$, $\delta_2=(\pi,-\pi)$, $\delta_3=(\pi,\pi)$, and $\delta_4=(-\pi,\pi)$, indicating  possible topological phase transitions.

\subsection{Details that extend beyond the dispersion relation}

\begin{figure}[htpb] 

    \begin{minipage}{0.5\textwidth}
        \includegraphics[width=0.45\linewidth,height=0.4\linewidth]{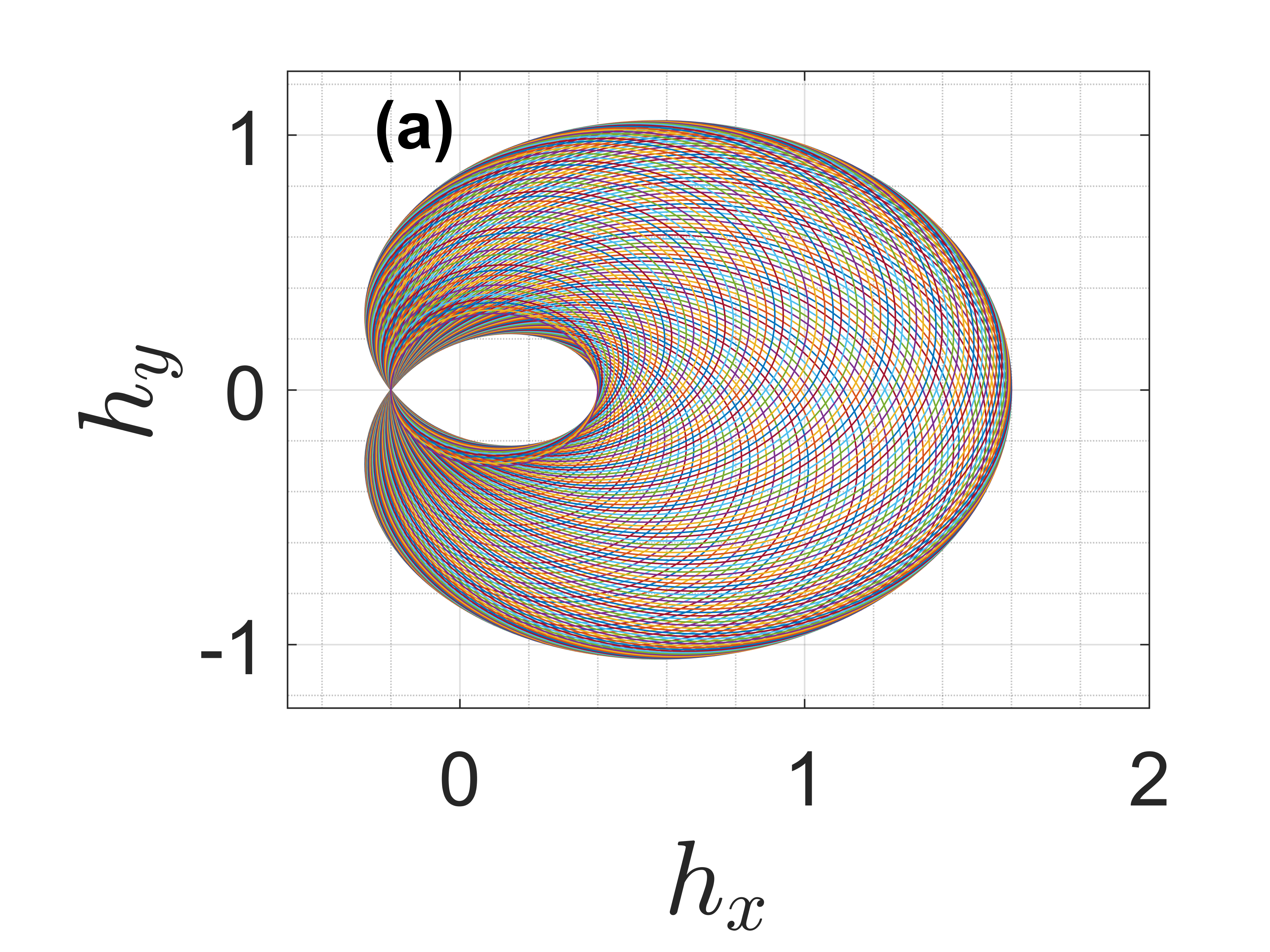}
        \hspace{10pt}
        \includegraphics[width=0.45\linewidth,height=0.4\linewidth]{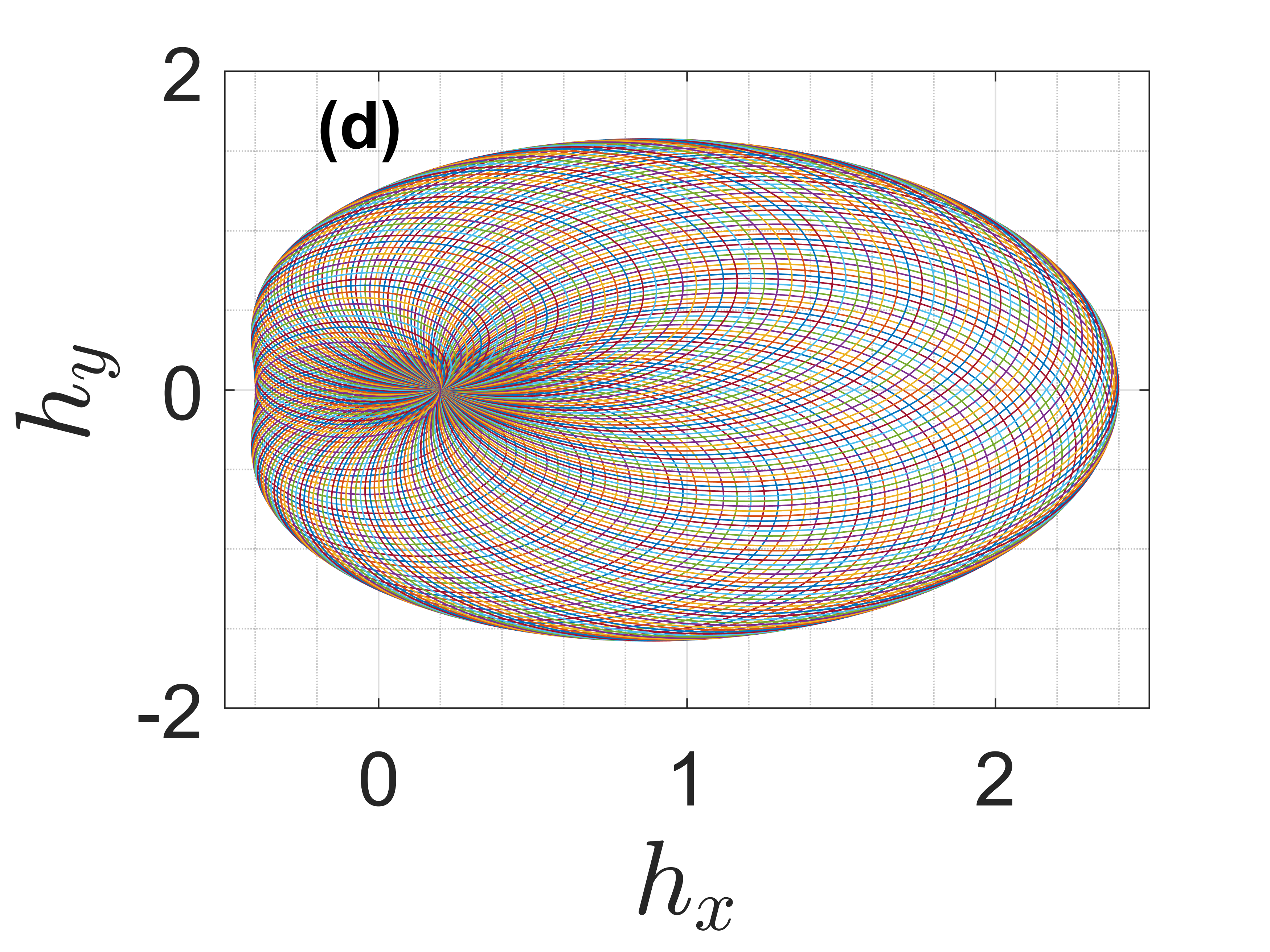}
\end{minipage}

\begin{minipage}{0.5\textwidth}
    \includegraphics[width=0.45\linewidth,height=0.4\linewidth]{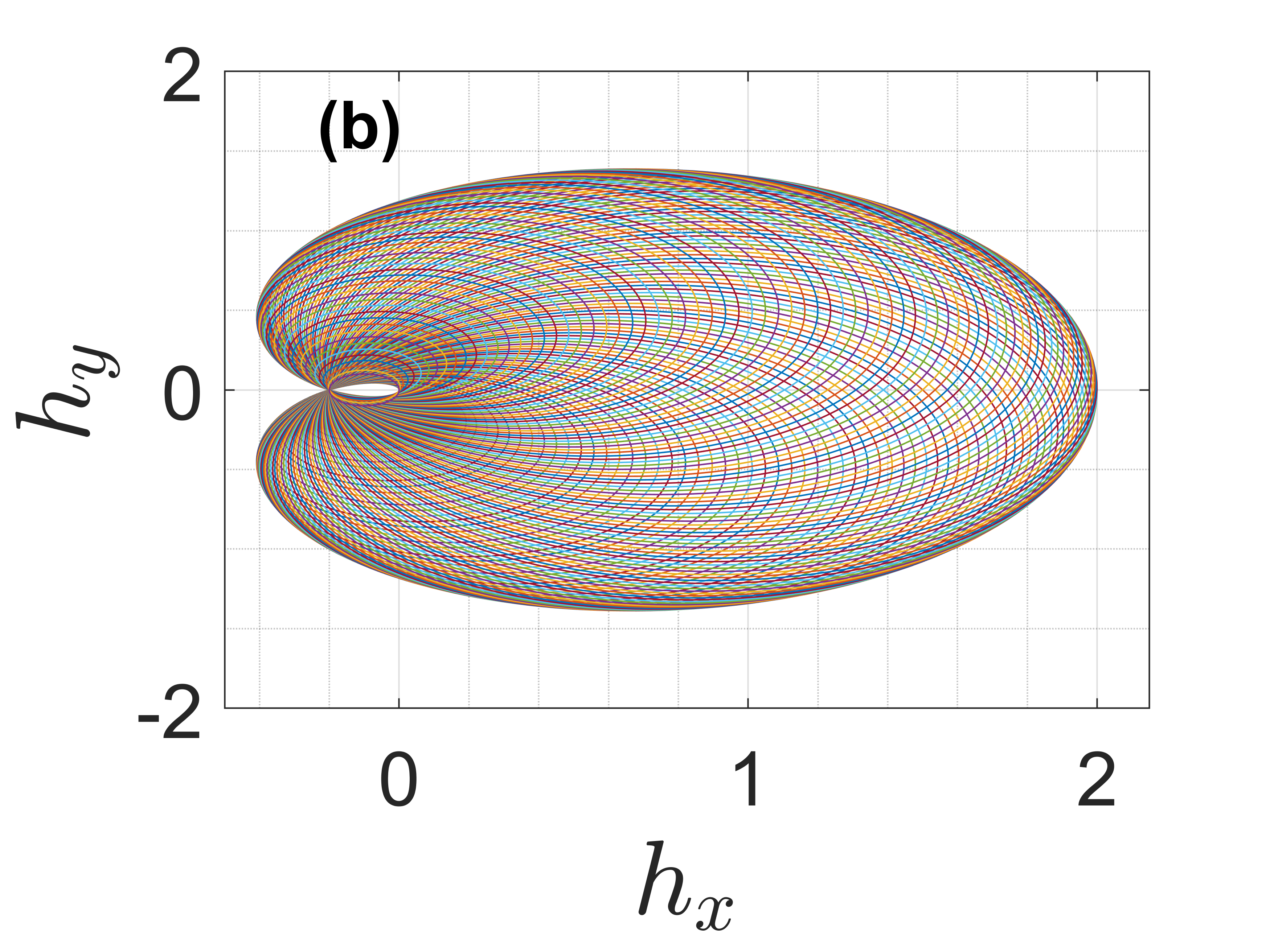}
        \hspace{10pt}
        \includegraphics[width=0.45\linewidth,height=0.4\linewidth]{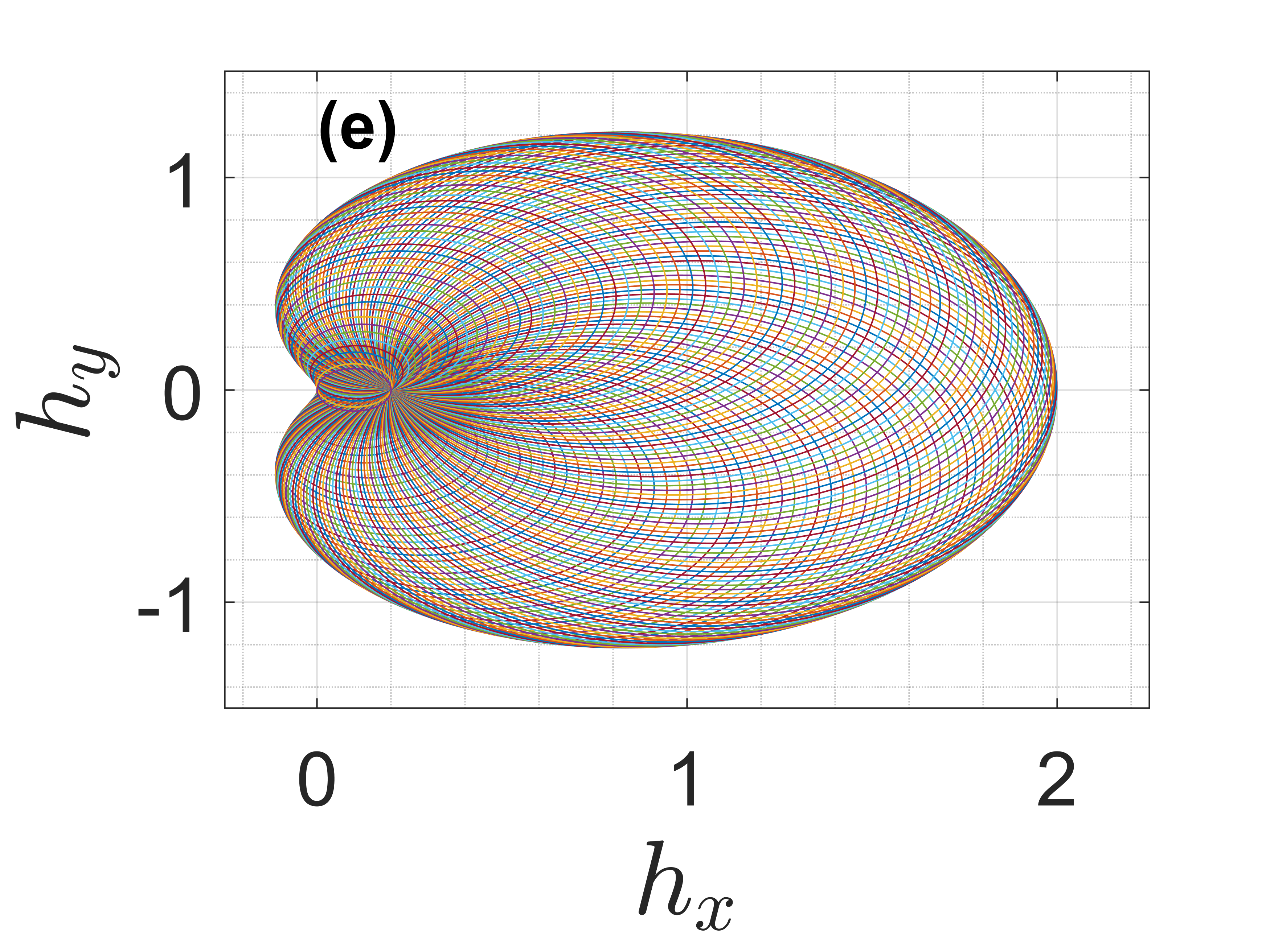}
\end{minipage}

\begin{minipage}{0.5\textwidth}
    \includegraphics[width=0.45\linewidth,height=0.4\linewidth]{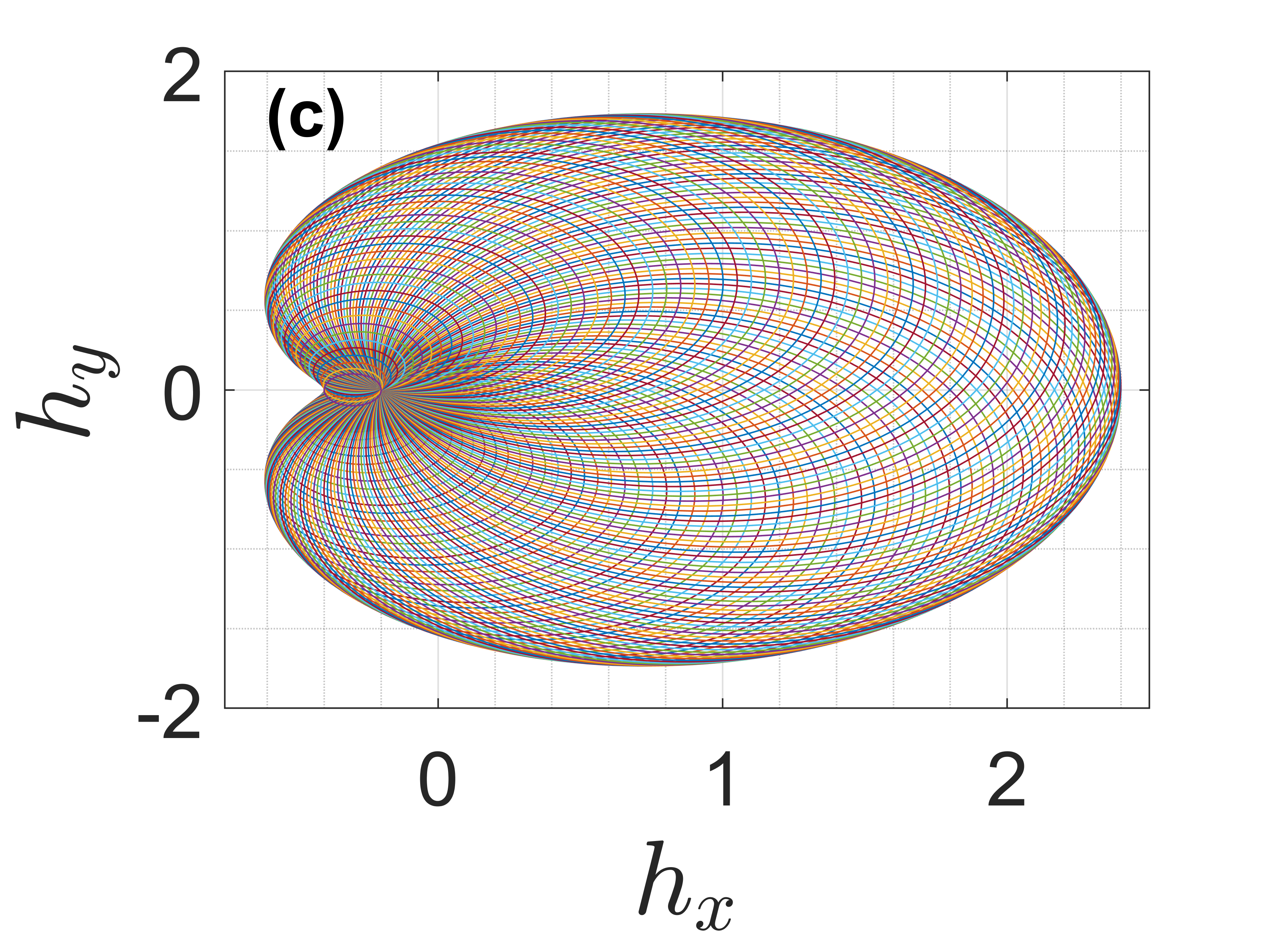}
    \hspace{10pt}
    \includegraphics[width=0.45\linewidth,height=0.4\linewidth]{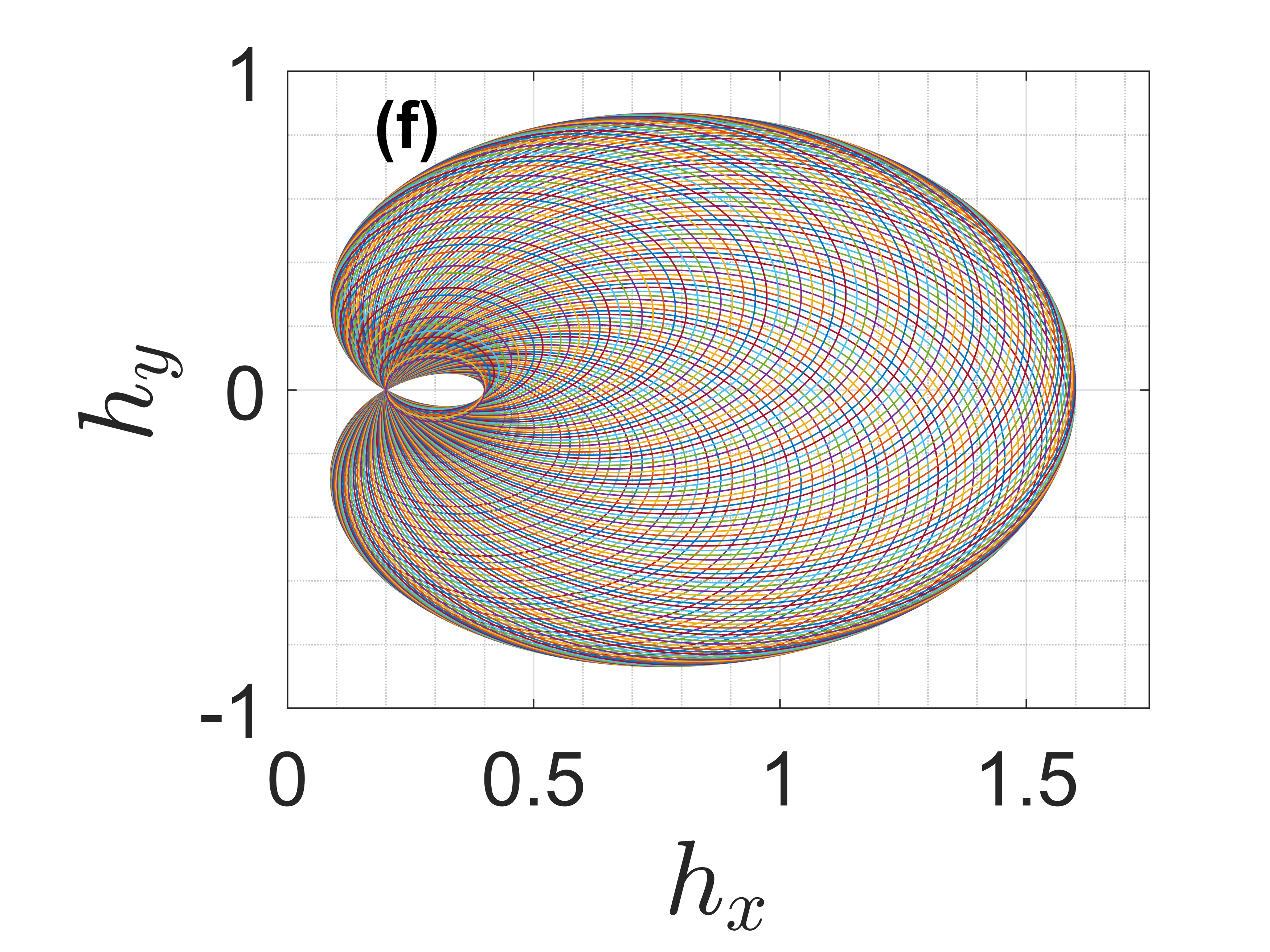}
\end{minipage}
    \caption{A group of closed curves for different scenarios is plotted in the 2D vector space  $(h_x,h_y)$. These closed curves are generated by choosing a group of paths created by meshing the parametric space $[k_x,k_y]$. Plots (a) to (c) represent the case when the individual topological atomic  chains are in a non-trivial topological phase $(v<w)$ with (a) $t<\frac{v+w}{2}$, (b) $t=\frac{v+w}{2}$ and (c) $t>\frac{v+w}{2}$. Plots (d) to (f) represent the case when the individual topological atomic  chains are in a trivial topological phase $(v>w)$ with (d) $t>\frac{v+w}{2}$, (e) $t=\frac{v+w}{2}$ and (f) $t<\frac{v+w}{2}$.}
    \label{fig:3}
\end{figure}

Based on the energy dispersion relation, the two band closing phenomenons unveiled a higher-order behavior, which is the signature of second-order topological transitions. In order to gain more information about the bulk and classify the topological nature of the system, we can look beyond this dispersion relation. Any hermitian $H(\mathbf{k})$ two-state bulk momentum space Hamiltonian can be decomposed using a linear combination of Pauli matrices.

\begin{equation}\label{eq:7}
    H(\mathbf{k})=h_{0}(\mathbf{k})\cdot I + \mathbf{h}(\mathbf{k})\cdot \boldsymbol{\sigma},
\end{equation}
where $h_{x}(k_x,k_y)$ and $h_{x}(k_x,k_y)$ are given in equation \ref{eq:6} and $\boldsymbol{\sigma}=(\sigma_x,\sigma_y,\sigma_z)$ are Pauli's matrices. The Hamiltonian at low energies $\mathbf{k}+\mathbf{\delta k}$ with $
\left|\mathbf{k}\right|\gg\left|\mathbf{\delta k}\right|$ behaves like massless Dirac fermions and has a linearized form as follows:
\begin{equation}\label{eq:8}
    H= (v-2t+w)\sigma_x + (w-t)(\delta k_x+ \delta k_y) \sigma_y.
\end{equation}

The band touching occurs when $t=\frac{v+w}{2}$, and the low-energy Hamiltonian takes the form $H=(w-t)(\delta k_x+ \delta k_y)\sigma_y$. So the sign of $(\delta k_x + \delta k_y)$ will decide the sign of the Fermi velocity. These low-energy excitations can lead to non-trivial transport properties in the system which we leave for future studies. The traditional technique for 2D systems is to use Berry curvature to define a topological invariant. However, the Chern number is not an appropriate topological invariant because the Berry curvature is zero.  Nonetheless, topology can emerge in lower dimensions by embedding 1D loops (e.g., non-contractible paths on the 2D BZ torus) that characterize the Berry phase, even when the 2D Chern number is zero.

To see how these different 1D closed loops capture the topology of the system, a group of closed curves are generated in the space created by the coefficients of Pauli matrices $(h_x,h_y)$ for each of the closed loops in BZ. To do that a mesh grid of $[k_x,k_y]$ was used to plot $(h_x,h_y)$ for each of the six scenarios. Figure \ref{fig:3} clearly shows that the closed curves are distinct and capture global aspects, indicating the  topological essence in the system. To better understand this global quality, we changed our technique to examine the phase of the Bloch state shown in equation \ref{eq:5}, which is also fundamental for the Berry phase.

\begin{figure}[htpb]
    \begin{minipage}{0.5\textwidth}
        \includegraphics[width=0.45\linewidth,height=0.35\linewidth]{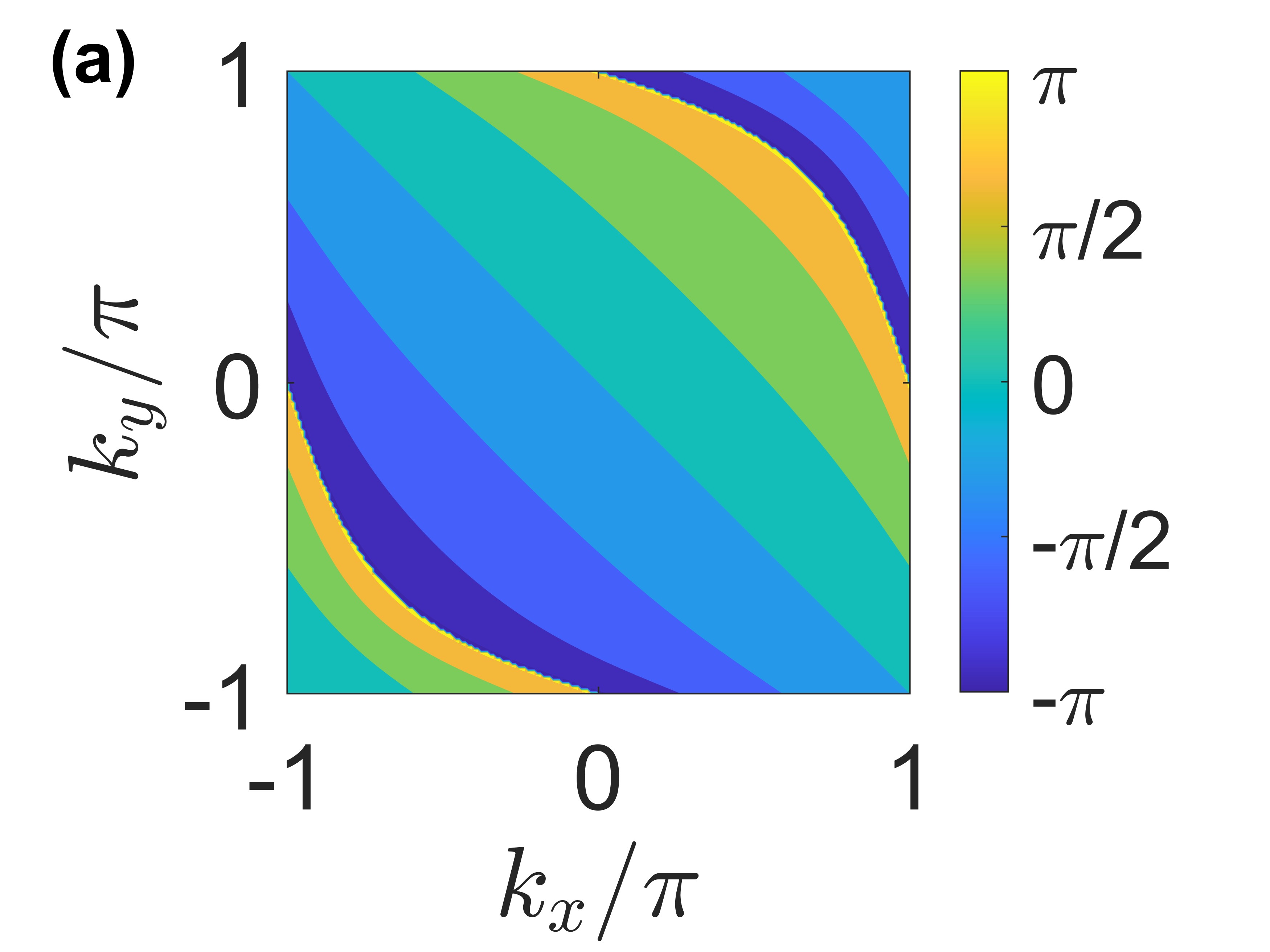} 
        \hspace{10pt}
        \includegraphics[width=0.45\linewidth,height=0.35\linewidth]{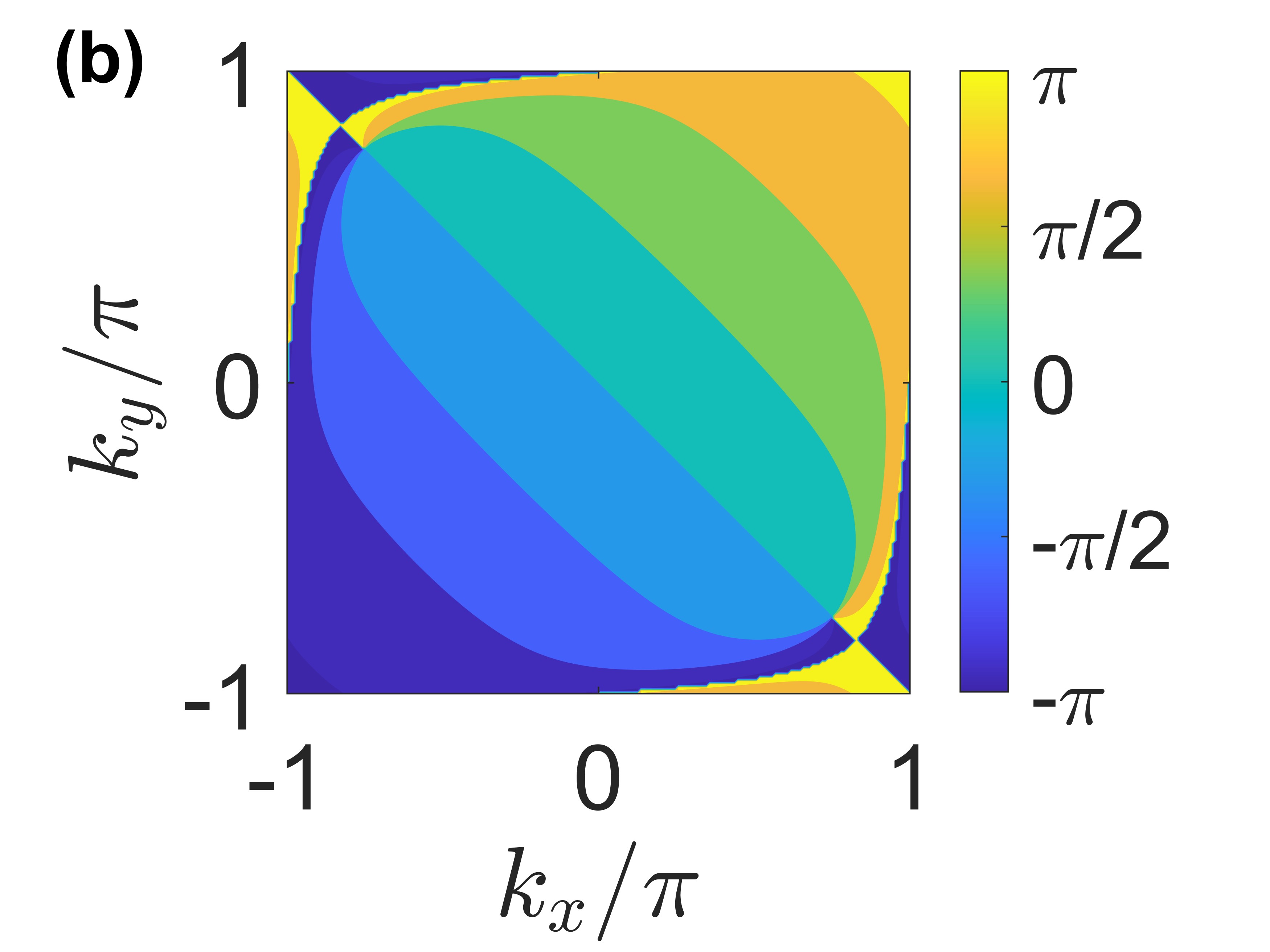}
    \end{minipage}

    \begin{minipage}{0.5\textwidth}
        \includegraphics[width=0.45\linewidth,height=0.35\linewidth]{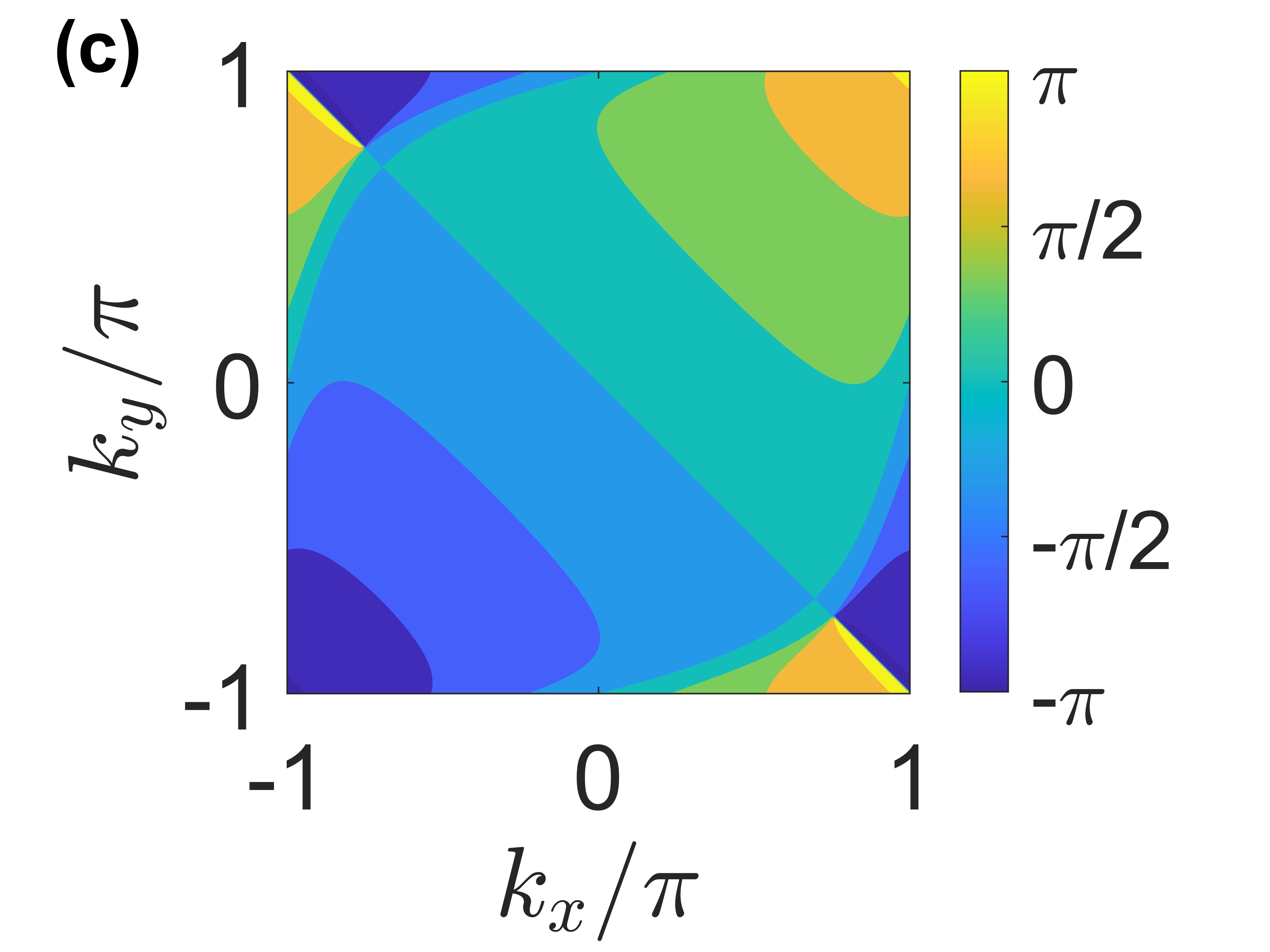} 
        \hspace{10pt}
         \includegraphics[width=0.45\linewidth,height=0.35\linewidth]{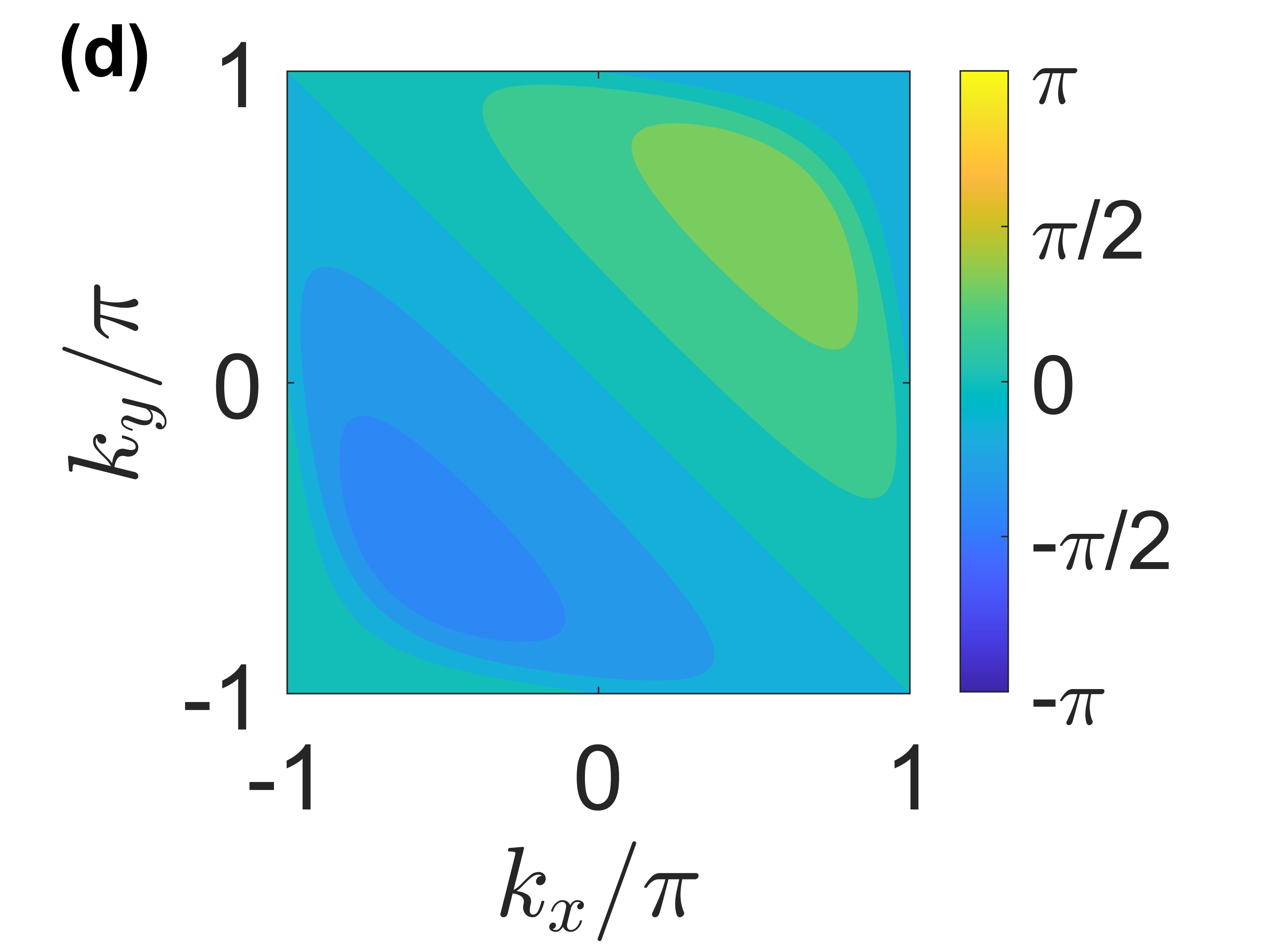}
    \end{minipage}
    \caption{Visual representation of the phase progression of the Bloch state in the form of color maps for a complete set of open and closed paths formed by using the meshing method in parametric space $[k_x,k_y]$. These color maps range from $[-\pi,\pi]$ where lemon color indicates $+\pi$ and navy blue indicates $-\pi$. Plots (a) and (b) represent the case when the individual topological atomic chains are in a non-trivial topological phase $(v<w)$ with $t<\frac{v+w}{2}$(a) and $t>\frac{v+w}{2}$(b). Plots (c) and (d) represent the case when the individual topological atomic chains are in a trivial topological phase $(v>w)$ with $t>\frac{v+w}{2}$(c) and $t<\frac{v+w}{2}$(d).}
    \label{fig:4}
\end{figure}

The phase progression that occurs during the process is shown in Figure \ref{fig:4} for four distinct settings.  This allows to identify the whole set of closed and open paths that a particle can take to get from one point to another in the BZ. 
Previously, four points were identified in parametric space (BZ), and we study the phase progression along the paths generated by joining any of those points, $\delta_1$, $\delta_2$, $\delta_3$, and $\delta_4$. Joining $\delta_1$ and $\delta_3$ or $\delta_2$ and $\delta_4$ creates a closed path. Pathways generated by $\delta_1$ and $\delta_3$ can contain $\delta_2$ or $\delta_4$, and vice versa for pathways created by $\delta_2$ and $\delta_4$. The diagonal line is a path that connects $\delta_2$ and $\delta_4$ with the constraints $-k_x=k_y$ and $k_x=-k_y$. Due to symmetric restrictions, $h_y=0$ along the diagonal path, resulting in $\theta(k_x,k_y)=0$, as seen in all four scenarios of phase progression. Any random path connecting $\delta_2$ and $\delta_4$, even if closed, does not account for an appropriate phase progression for the particle in the BZ. A phase progression of $4\pi$ for $v<w$ with $t<\frac{v+w}{2}$, $2\pi$ for both $v<w$ with $t>\frac{v+w}{2}$ and $v>w$ with $t>\frac{v+w}{2}$, and finally $0$ for $v>w$ with $t<\frac{v+w}{2}$ in the path generated by joining $\delta_1$ and $\delta_3$. Numerically, any arbitrary closed paths formed by connecting $\delta_1$ and $\delta_3$ points contribute to identifying the topological phases in the system. The winding number technique captures the mathematical justification for the above explanation and is calculated as follows:

\begin{figure}[htpb] 

    \begin{minipage}{0.5\textwidth}
        \includegraphics[width=0.45\linewidth,height=0.4\linewidth]{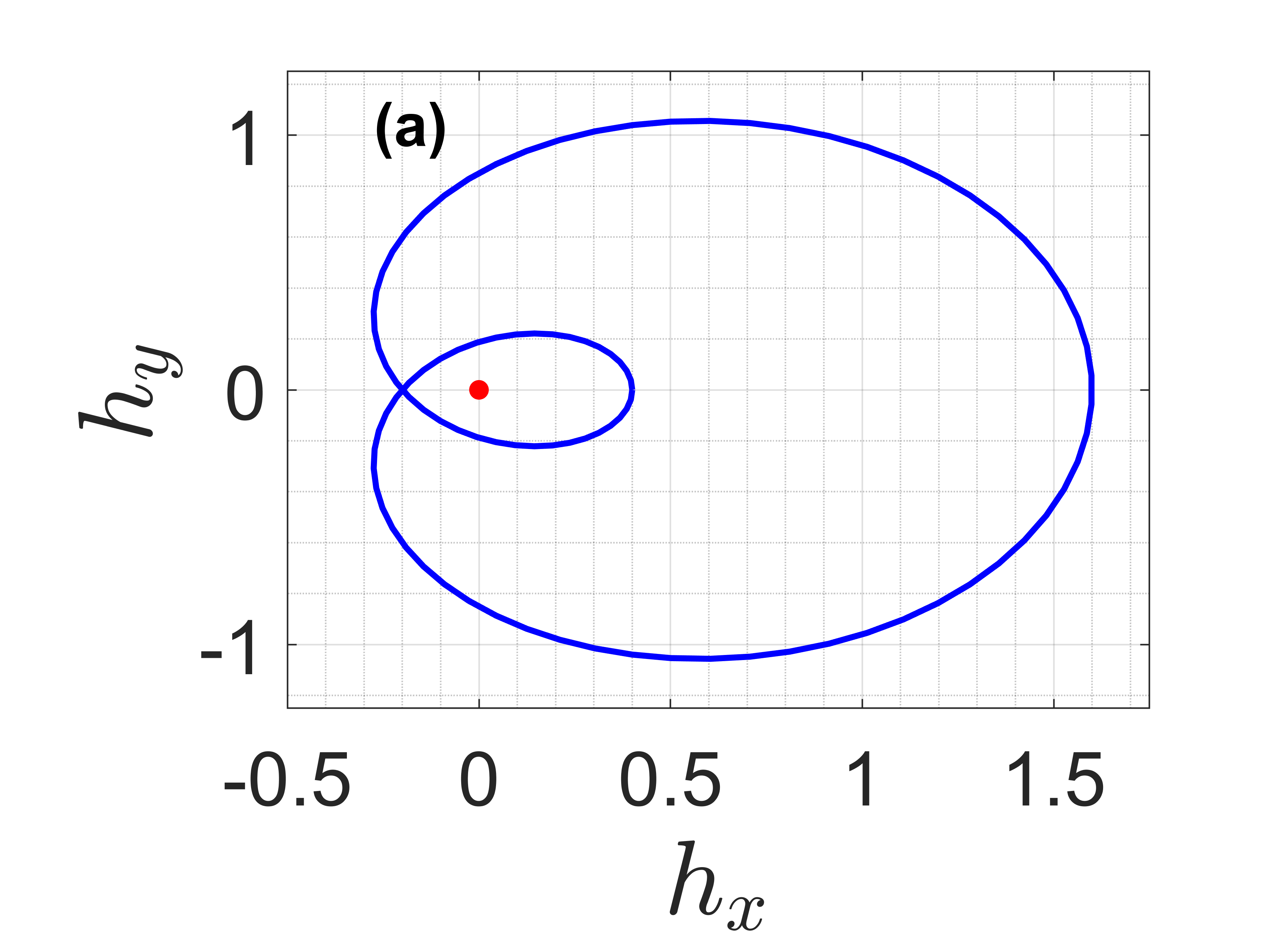}
        \hspace{10pt}
        \includegraphics[width=0.45\linewidth,height=0.4\linewidth]{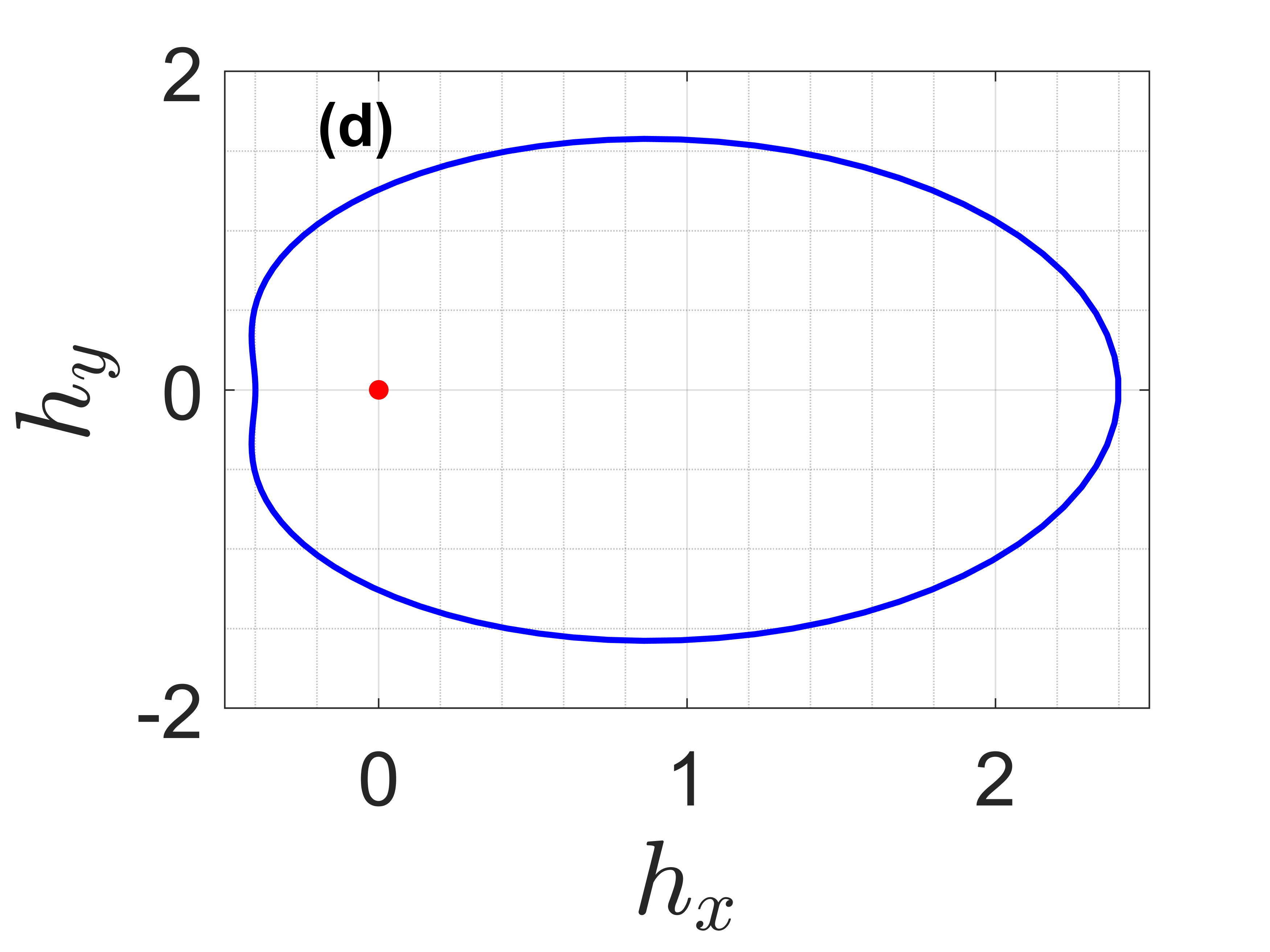}
\end{minipage}

\begin{minipage}{0.5\textwidth}
    \includegraphics[width=0.45\linewidth,height=0.4\linewidth]{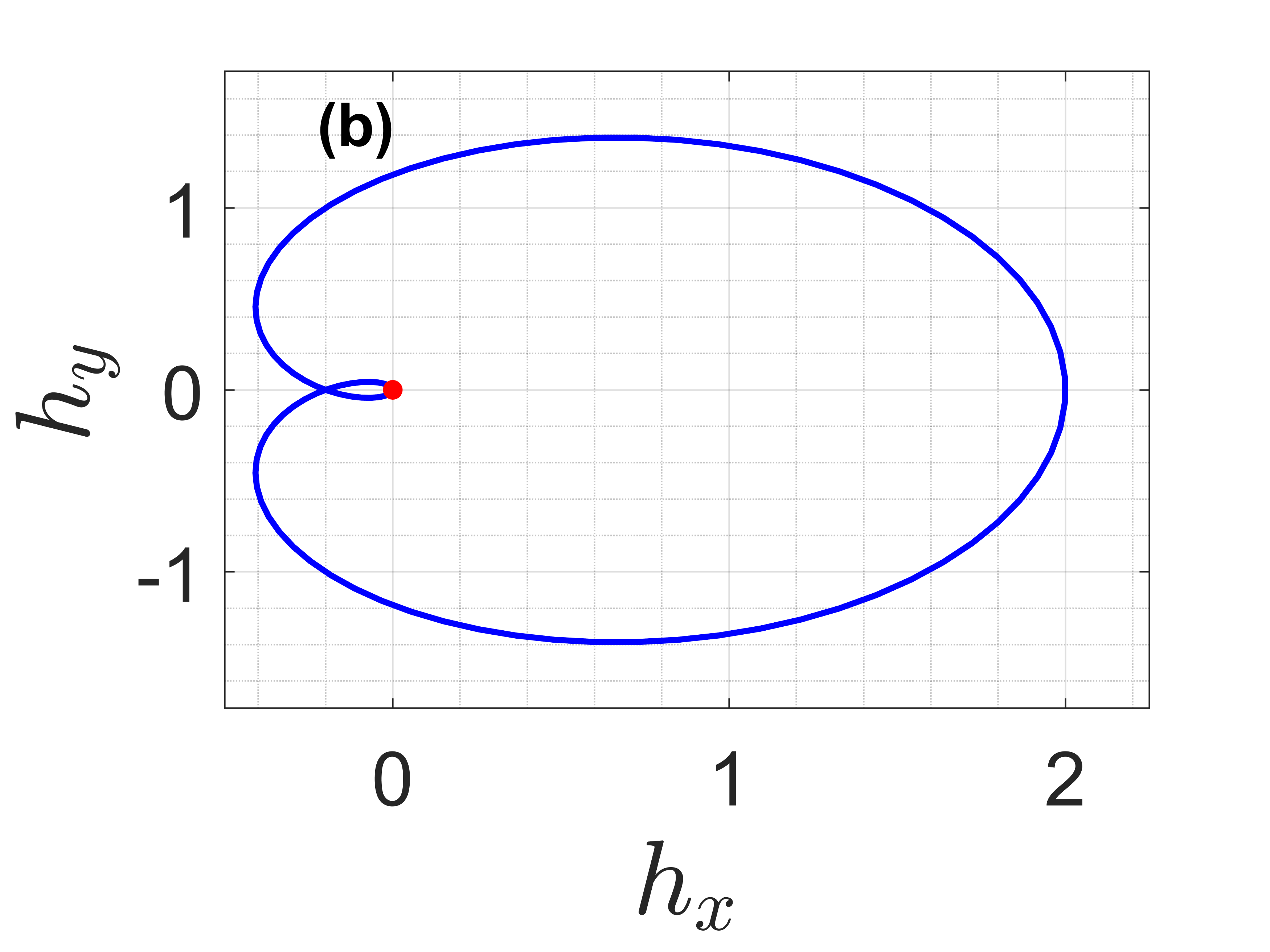}
        \hspace{10pt}
        \includegraphics[width=0.45\linewidth,height=0.4\linewidth]{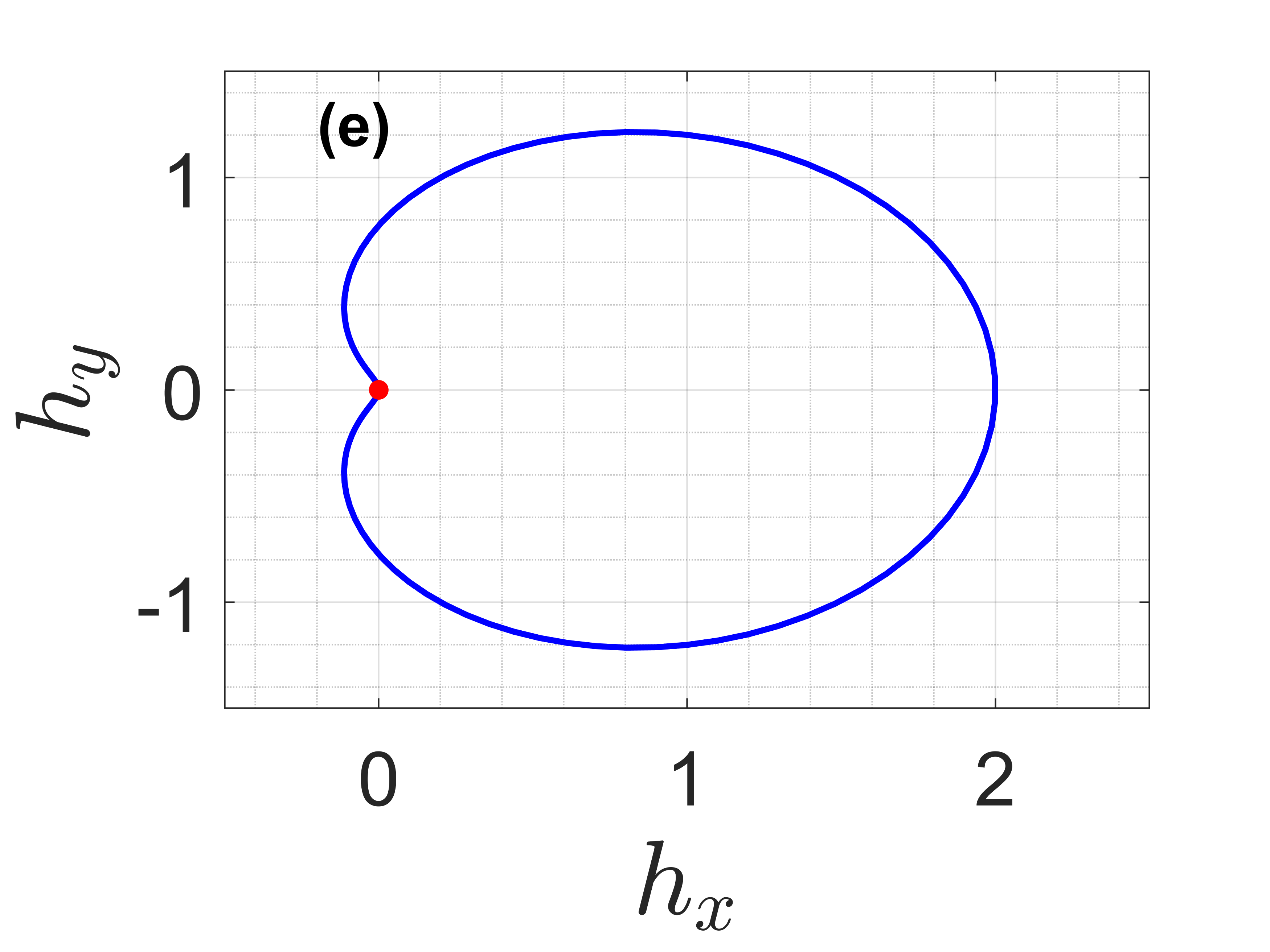}
\end{minipage}

\begin{minipage}{0.5\textwidth}
    \includegraphics[width=0.45\linewidth,height=0.4\linewidth]{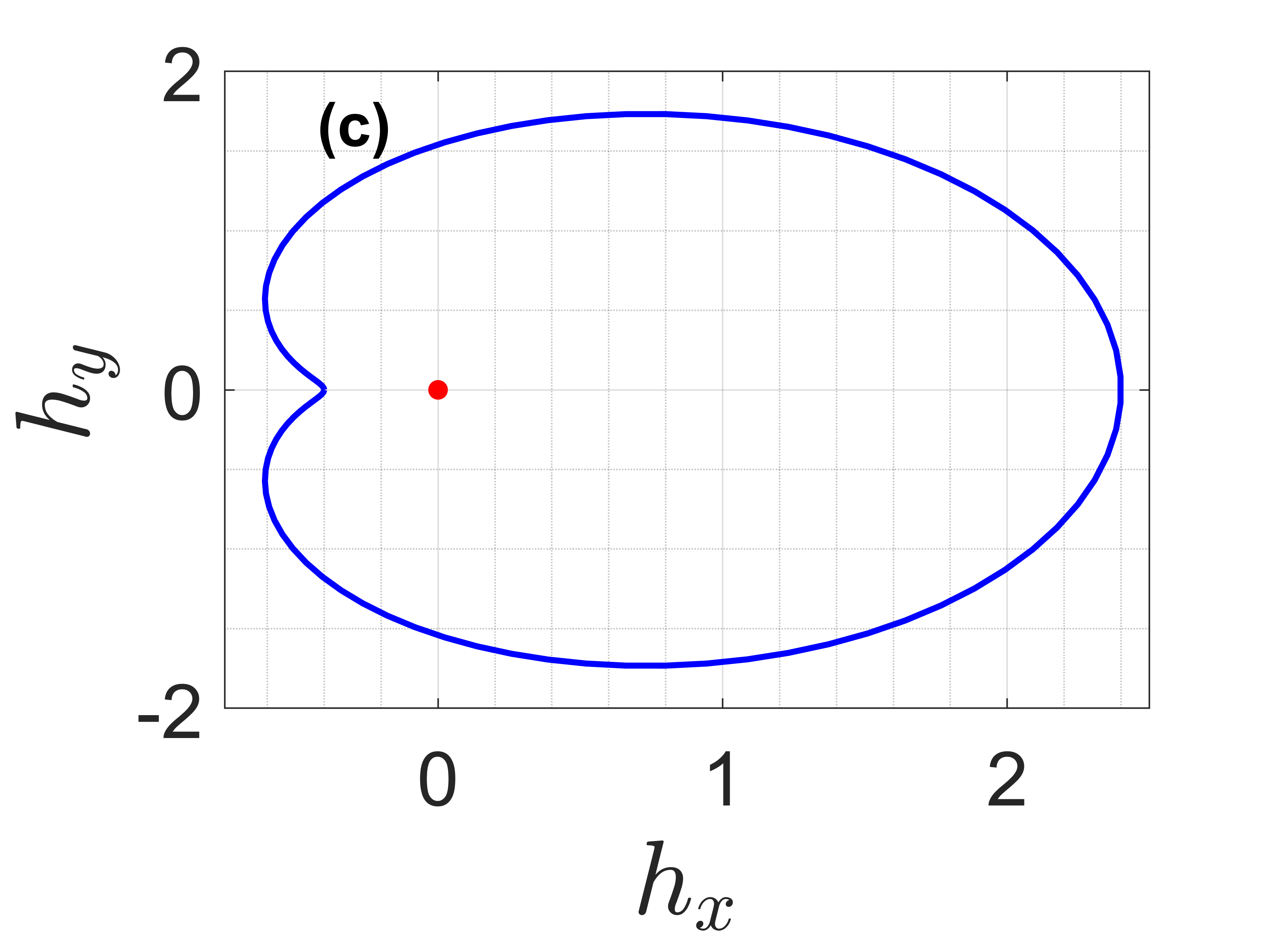}
    \hspace{10pt}
    \includegraphics[width=0.45\linewidth,height=0.4\linewidth]{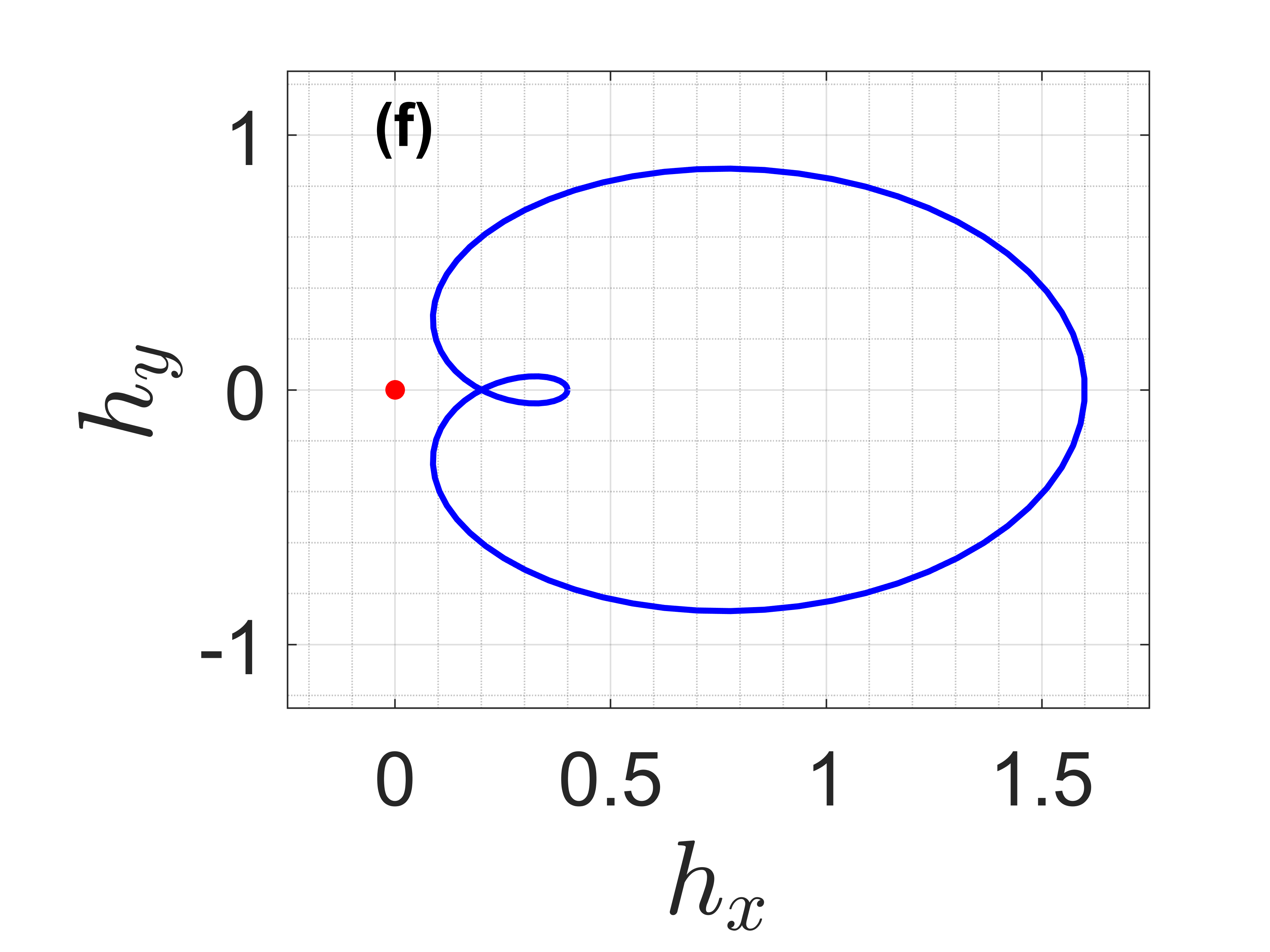}
\end{minipage}

    \caption{A schematic sketch of the closed curve for different scenarios are plotted in the 2D vector space  $(h_x,h_y)$ with the origin indicated as a red dot. These closed curves are generated by choosing a specific closed path in the parametric space $(k_x,k_y)$. The closed path satisfies the condition $k_x=k_y$. Plots (a) to (c) represent the case when the individual topological atomic chains are in a non-trivial topological phase $(v<w)$ with (a) $t<\frac{v+w}{2}$ with $\nu=2$, (b) $t=\frac{v+w}{2}$ with $\nu=$ undefined and (c) $t>\frac{v+w}{2}$ with $\nu=1$. Plots (d) to (f) represent the case when the individual topological atomic chains are in a trivial topological phase $(v>w)$ with (d) $t>\frac{v+w}{2}$ with $\nu=1$, (e) $t=\frac{v+w}{2}$ with $\nu=$ undefined and (f) $t<\frac{v+w}{2}$ with $\nu=0$.}
    \label{fig:5}
\end{figure}

\begin{equation}\label{eq:9}
    \nu =-\frac{1}{\pi}\int_{\delta_1}^{\delta_3} \mathbf{A}(\mathbf{k})\cdot\mathbf{dk}=
    \begin{cases} 
    2, & v<w; t<\frac{v+w}{2}, \\
    
     1, & v<w; t>\frac{v+w}{2}, \\
     1, & v>w; t>\frac{v+w}{2}, \\
   
    0, &  v>w; t<\frac{v+w}{2}. \\
    \end{cases}
\end{equation}

Figure \ref{fig:5} depicts the winding of a loop around the origin for the path $k_x=k_y$. This path is reliable for an appropriate topological invariant to classify the system's topology. At the origin, the vector $\mathbf{h(k)}$ has an infinite direction and zero magnitude, causing an indefinite impact on the phase of the Bloch state. A phase transition corresponds with the origin, and loops encompassing it are topologically equal and distinct from other loops. A non-trivial topological phase happens when the curve wraps around the point once or twice, whereas a trivial topological phase occurs when the winding is zero. Depending on the pre-factor, the Zak phase can also distinguish the topological phases.\\
\begin{figure}[htpb]
    \includegraphics[width=0.7\linewidth,height=0.55\linewidth]{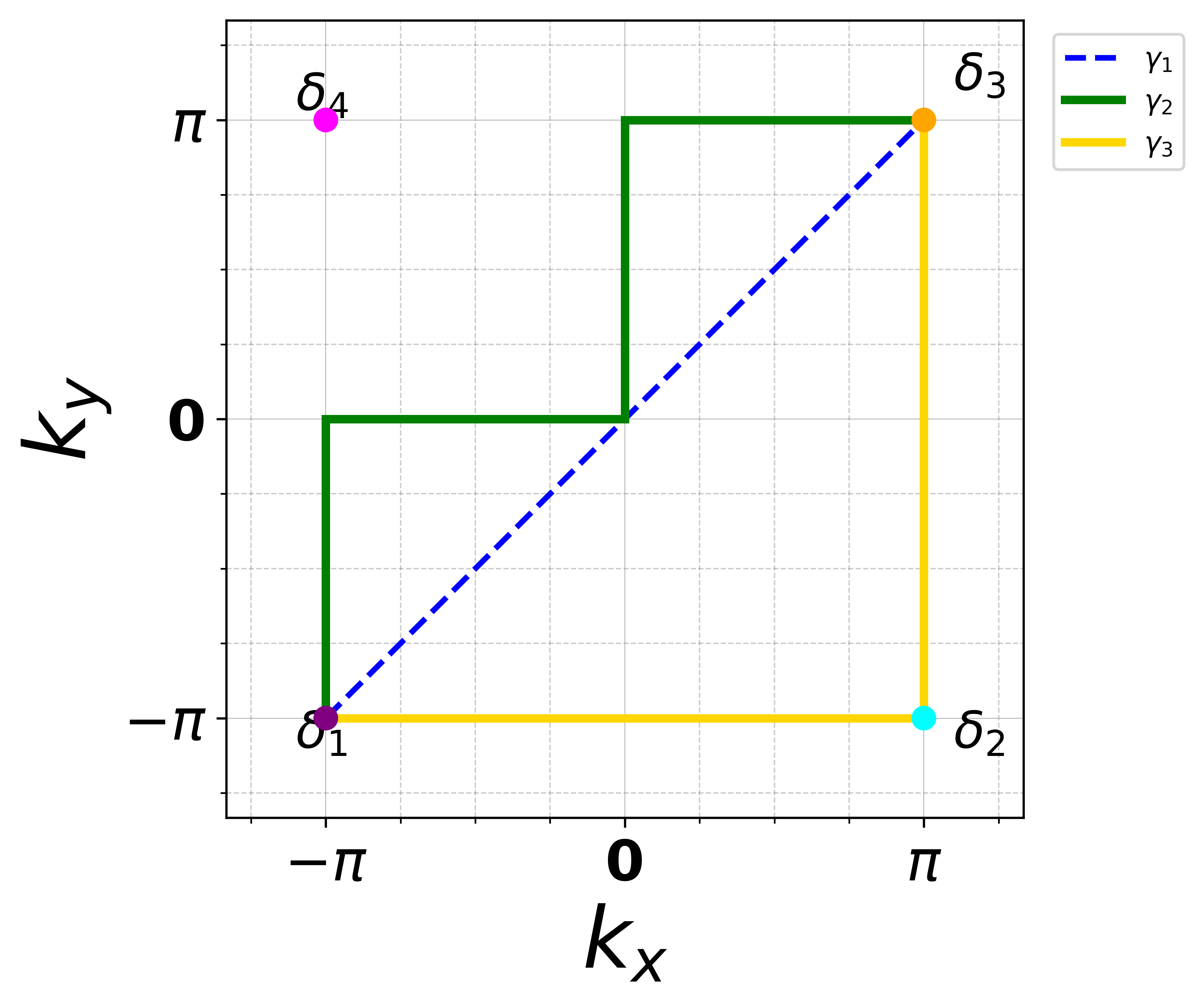} 
    \caption{A sketch to  distinguish between the  closed path connecting the points $\delta_1=[-\pi,-\pi]$ and $\delta_3=[\pi,\pi]$ in the parametric space which is a torus. The two closed paths $\gamma_1$ (blue dash line) and $\gamma_2$ (green line), connecting $\delta_1$ and $\delta_3$, are closed loops, which can be deformed continuously into each other on the torus. Where as,  $\gamma_3$ (yellow line),  which includes $\delta_2$ in the pathway, behaves as two closed loops orthogonal to each other, intersecting at $\delta_2$.  Here, $\gamma_1$ and $\gamma_2$ are homotopic curves in the torus, where as,  $\gamma_3$ is not homotopic to $\gamma_1$.  }
    \label{fig:6}
\end{figure}

The other two points $\delta_2$ or $\delta_4$ can also be included in the pathway connected by $\delta_1$ and $\delta_3$, as shown in Figure \ref{fig:6}. We choose two curves $\gamma_1$ and $\gamma_3$ (which includes $\delta_2$ in the pathway) in the parametric space $T=[-\pi,\pi]\times[-\pi,\pi]$, with $\delta_1$ as the starting point and $\delta_3$ as ending point for both the curves. We noticed that $\gamma_1$ and $\gamma_3$ do not belong to the same homotopy group; in a sense, it is impossible to deform $\gamma_1$ into $\gamma_3$ continuously. To explain it in detail, consider $\gamma_3$ as shown in Figure \ref{fig:6} and the path traversed by this curve is like 
two perpendicular closed paths (circles) intersecting each other at a point $\delta_2$ in the parametric space T. While as, both $\gamma_1$ (and $\gamma_2$) are single continuous closed path that avoid such intersection in T. So $\gamma_1$ cannot be continuously deformed into $\gamma_3$, without introducing a discontinuity which requires a cutting at intersection point and reattaching the four ends to form a big closed loop, which is not allowed in the framework of homotopy of curves. Although both curves start and end at the same points, we only choose an arbitrary path that can connect $\delta_1$ and $\delta_3$ without $\delta_2$ or $\delta_4$ points in the pathways. Those pathways are suitable for calculating winding number to distinguish the topological phases in the system. For computational ease, $\gamma_1$ is considered for analytical calculation in equation \ref{eq:9}.

\subsection{Bulk Polarization}

 The modern theory of polarization \cite{PhysRevB.47.1651,Spaldin2012,PhysRevLett.120.026801} describes the bulk polarization in terms of the Berry phase. The same Berry phase is used to capture the bulk topological invariant from the previous section. Given the model's non-trivial stacking in the diagonal direction with crystalline mirror symmetry, we defined polarization along the closed loops, which were mentioned previously, to derive the bulk topological invariant. Any closed loop which belongs to the same homotopy class of $\gamma_1$ in the momentum space gives the same result. For analytical ease and experimental feasibility, we choose the line $k_x=k_y$ direction and define diagonal polarization as follows:

\begin{equation}
    \boldsymbol{P}_{xy} = \frac{e}{2\pi}\int_{Bz} \mathbf{A}(\mathbf{k})\cdot\mathbf{dk}=
    \begin{cases} 
    2e, & v<w; t<\frac{v+w}{2}, \\
     1e, & v<w; t>\frac{v+w}{2}, \\
     1e, & v>w; t>\frac{v+w}{2}, \\
    0, &  v>w; t<\frac{v+w}{2}. \\
    \end{cases}
\end{equation}

\begin{figure}[htpb]
    \begin{minipage}{0.5\textwidth}
        \includegraphics[width=0.45\linewidth,height=0.35\linewidth]{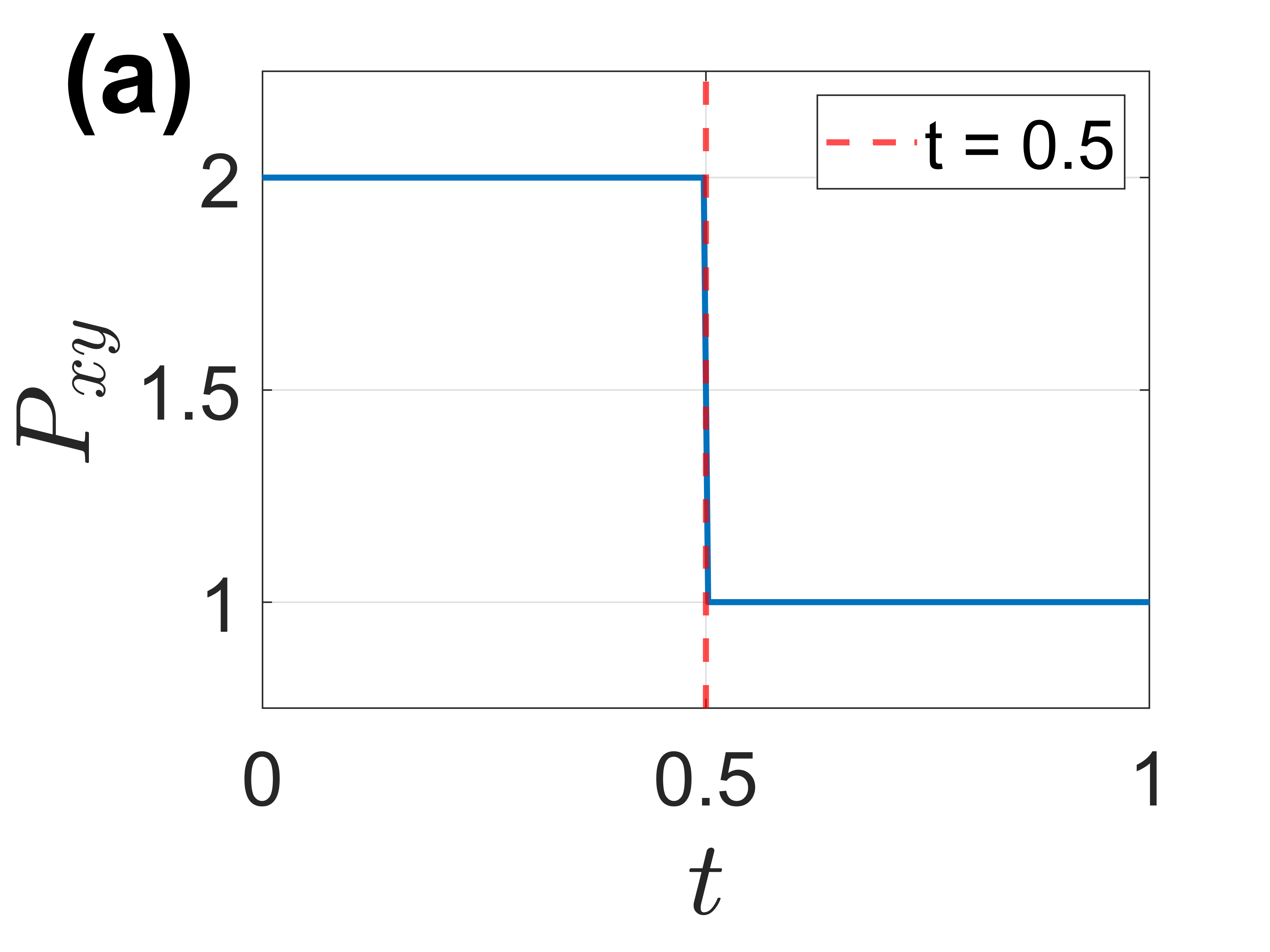} 
        \hspace{10pt}
        \includegraphics[width=0.45\linewidth,height=0.35\linewidth]{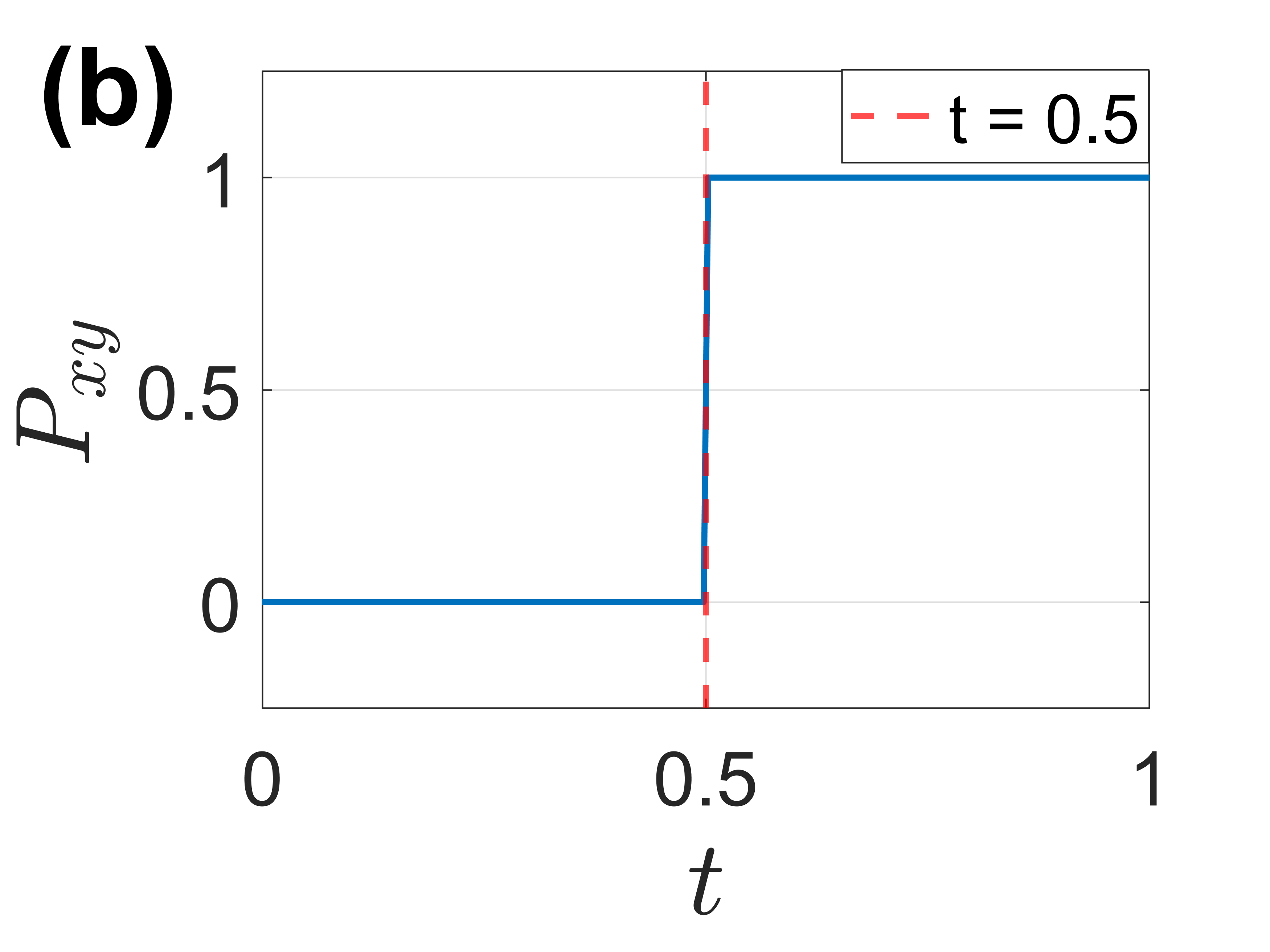}
    \end{minipage}
    \caption{The diagonal bulk polarization $P_{xy}$ is plotted as a function of inter-chain hopping $t$. When the individual chains are in a non-trivial topological phase $(v<w)$, the bulk polarization quantize to 4 for $t<\frac{v+w}{2}$ and to 2 for $t>\frac{v+w}{2}$, as shown in the left plot. When the individual chains are in a trivial topological phase $(v>w)$, the bulk polarization quantize to 2 for $t>\frac{v+w}{2}$ and vanishes for $t<\frac{v+w}{2}$, as shown in the right plot.}
    \label{fig:7}
\end{figure}

According to Figure \ref{fig:7}, when the individual chains are in a non-trivial topological phase, the bulk polarization quantizes to 2 when $t<\frac{v+w}{2}$, to 1 when $t>\frac{v+w}{2}$, and a leap in the value when $t=0.5$. When the chains are in a trivial topological phase $(v<w)$, the bulk polarization quantizes to 1 when $t>\frac{v+w}{2}$, to 0 when $t<\frac{v+w}{2}$, and a leap occurs when $t=0.5$. The change in polarization is defined up to modulo of the elementary charge e. Because of periodic boundary conditions on the lattice, any integer multiple of charge pumping across a unit cell implies no net change in the polarization. For example, if the polarization along a particular direction is $1e$, the change in polarization in that direction is zero, but physically it represents a charge of one electron was pumped across the unit cell in that direction. Similarly, a diagonal polarization of $P_{xy}=2e$ indicates that two electrons are pumped per unit cell along the line $y=x$ direction. This shift is topologically trivial in the bulk in the sense that it is equal to zero because of modulo e.\\
But its consequences become crucial in a finite system. In a system with finite boundaries, the charges that are pumped have no unit cell at the boundary to accommodate them. Hence they accumulate there as boundary charges of the system. For higher-order topological insulators in 2D accumulations of charges are seen at the corners of the system rather than the edges. Therefore, a bulk polarization of $2e$ along the diagonal strongly suggests that two electrons worth of charge can be seen at the system's boundary. This is consistent with the expected localized corner zero modes in our model, as discussed in Section \ref{sec: Bulk-Boundary Correspondace}, which illustrates the bulk-boundary correspondence for second-order topological phase.

Since the model can also be seen as the stacking of effective topological atomic chains with different parameters in the y-direction or x-direction. Calculating the polarization in individual directions can reveal further information about the topological nature of the system. The change in polarization  is represented as $\delta\boldsymbol{P}=(\delta P_{x}, \delta P_{y})=(0.5,0.5)$. It causes charges to shift to the corners of the system. The crystalline mirror symmetry along the $y=x$ line  restricts the accumulation of charges to symmetric corners. This is in contrast to uniformly accumulation at four corners observed in conventional two-dimensional stacking. The two corners at $x=-y$ and $-x=y$ lines are not protected by crystalline mirror symmetry. Therefore, charges are pumped to the corners of the lattice lying on the $y=x$ line.

\subsection{Symmetries of the system}

The unconventional stacking of topological atomic chains preserves time-reversal symmetry (TRS), inversion symmetry (IS), chiral symmetry (CS) and particle-hole symmetry (PHS),  which are intrinsic symmetries of the 1D topological atomic chains. The presence of these symmetries imposes certain conditions on the Bloch Hamiltonian $H(\boldsymbol{K})$ in $\boldsymbol{k}$-space as follows:
\begin{center}
        $\mathcal{T}H(k_x,k_y)\mathcal{T}^{-1}=H(-k_x,-k_y),$\\
        $\mathcal{I}H(k_x,k_y)\mathcal{I}^{-1}=H(-k_x,-k_y),$\\
        $\mathcal{C}H(k_x,k_y)\mathcal{C}^{-1}=-H(k_x,k_y),$\\
        $\mathcal{P}H(k_x,k_y)\mathcal{P}^{-1}=-H(-k_x,-k_y).$
\end{center}
where, $\mathcal{T}$, $\mathcal{I}$, $\mathcal{C}$ and $\mathcal{P}$ represent the TRS, IS, CS and PHS operators, respectively. Our model defines these symmetries as $\mathcal{T}=\mathcal{K}$, where $\mathcal{K}$ is a complex conjugation, $\mathcal{I}=\sigma_{x}$, $\mathcal{C}=\sigma_{z}$ and $\mathcal{P}=\mathcal{C}\mathcal{T}$, where $\sigma_{x}$ and $\sigma_{z}$ are Pauli matrices.  \\

The coexistence of TRS and IS reinforces vanishing Berry curvature in the BZ resulting in zero Chern number. Therefore, system appears to have trivial topological nature in the conventional sense, but it may host non-trivial  higher-order topological phases. Within the parametric space, a closed trajectory along the diagonal direction connecting $\delta_2$ and $\delta_4$ fails a non-trivial mapping to the vector space. This is because TRS constrains the $h_y$ component to zero, effectively reducing the  vector space to be one-dimensional manifold. IS facilitates band inversion, leading to band crossing where valence and conduction bands exchange their characters, as $(k_x,k_y)$ varies. This results in a topologically protected state at the boundary. In our model, the orientation of the topological atomic chains along the line $y=x$ introduces additional symmetry in the form of crystalline mirror symmetry. The mirror symmetry $\mathcal{M}_{y=x}$ projects the system onto itself when reflected across $y=x$, imposing limitations on form and location of  boundary states. This symmetry enforces the zero-energy modes localized exclusively at the corners along the mirror-symmetric line, resulting in the emergence of corner states, indicating the possibility of higher-order topological phases.  These features are discussed further  in Section \ref{sec: Bulk-Boundary Correspondace}.

\section{The edge of the system}
\label{sec: The edge of the system}
\begin{figure}[htpb]
    \begin{minipage}{0.55\textwidth}
         \includegraphics[width=0.35\linewidth,height=0.3\linewidth]{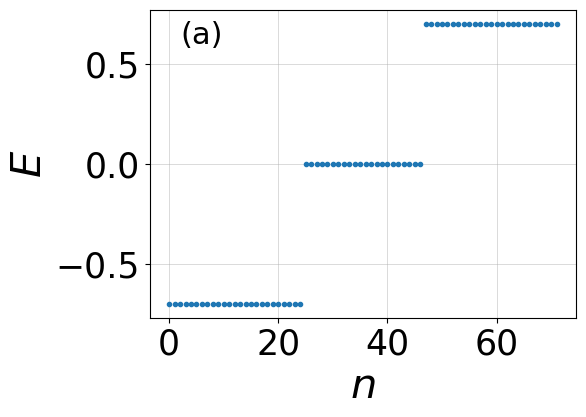} 
        \hspace{5pt}
        \includegraphics[width=0.35\linewidth,height=0.3\linewidth]{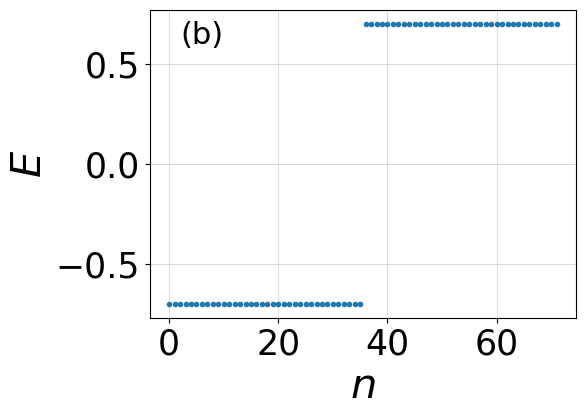}
    \end{minipage}
    \caption{The energy spectrum of the 2D extension topological atomic  model's Hamiltonian when the system is in its finite size by OBC. The energy spectrum plotted on (a) is when the individual topological atomic  chains are in one of the extreme cases, where $v=0$ and  $w \neq 0$ with $t=0$ . The energy spectrum plotted on (b) is when the individual topological atomic  chains are in its other extreme cases, where $w=0$ and $v \neq 0$ with  $t=0$. Zero modes appear in the energy spectrum when the individual chains are in a non-trivial topological phase.  }
    \label{fig:8}
\end{figure}

In the previous section, we emphasized a CPBC and calculated the bulk topological invariant. In this section, we attempt to comprehend the system's behavior in its finite state. Because the system is in 2D, we analyze it under two scenarios: partial periodic boundary condition (PPBC) and open boundary condition (OBC). In PPBC, we can impose periodic boundary conditions in one direction while leaving the other open, resulting in a cylindrical configuration with the cylinder's axis parallel to the PPBC direction. The following are the two Hamiltonians in the PPBC. 

 \begin{equation}
    \begin{aligned}
      \hat{H}_{CYL}(k_x)&= \sum_{m_y=1}^{N_y} (v+te^{ik_x})\ket{m_y,A}\bra{m_y,B}\\&
     +\sum_{m_y}^{N_y-1} (t+we^{ik_x})\ket{(m_y+1),A}\bra{m_y,B}+h.c.,  
    \end{aligned}
\end{equation}

\begin{equation}
    \begin{aligned}
      \hat{H}_{CYL}(k_y)&= \sum_{m_x=1}^{N_x} (v+te^{ik_y})\ket{m_x,A}\bra{m_x,B}\\&
     +\sum_{m_x}^{N_x-1} (t+we^{ik_y})\ket{(m_x+1),A}\bra{m_x,B}+h.c.,
    \end{aligned}
\end{equation}
where $\hat{H}_{CYL}(k_x)$ depicts a model of N stacked topological atomic  chains glued to the surface of the cylinder, whose axis is parallel to the x-direction. $\hat{H}_{CYL}(k_y)$ is a similar model in the y-direction. The equations above resemble those of a 1D topological atomic  chain, with inter-hopping as $v+te^{ik_x(k_y)}$ and intra-hopping as $t+we^{ik_x(k_y)}$. The added parameter $k_x(k_y)$ controls hopping strengths, resulting in a 1D topological atomic  chain for every value of $k_x(k_y)$, with the constraint $v_0=(v+te^{ik_x(k_y)})<w_0=(t+we^{ik_x(k_y)})$ for a non-trivial topological phase. The related invariant is as follows:
\begin{equation}
    \nu =
    \begin{cases} 
    1, & |v_0| < |w_0|, \\
    0, &  |v_0| > |w_0|. \\
    \end{cases}
\end{equation}
Reference \cite{47.Agrawal_2023} provides a framework for further exploration of such systems. We examine the OBC on the system to comprehend boundary effects and distinguish non-trivial and trivial topological phases. In extreme cases, diagonalizing the Hamiltonian of equation \ref{eq:1} produces the spectrum shown in Figure \ref{fig:8}. The appearance of zero modes when the system is in its non-trivial phase depicts the topological nature of the system.

However, to cross-check the boundary effects of the system, we again considered the CPBC, but this time, we diagonalized the Hamiltonian in the real space rather than in the momentum space. The Hamiltonian of the system when CPBC in real space is as follows:
 \begin{equation}
        \begin{aligned}
          \hat{H}& = \sum_{m_1}^{N_x}\sum_{m_2}^{N_y} v\ket{m_1,m_2,A}\bra{m_1,m_2,B}\\ &+\sum_{m_1}^{N_x}\sum_{m_2}^{N_y} w\ket{(m_1 \text{mod} N_x),(m_2 \text{mod} N_y),A}\bra{m_1,m_2,B} \\
          &+\sum_{m_1}^{N_x}\sum_{m_2}^{N_y} t\ket{(m_1 \text{mod} N_x),m_2,A}\bra{m_1,m_2,B} \\
           &+\sum_{m_1}^{N_x}\sum_{m_2}^{N_y} t\ket{m_1,(m_2 \text{mod} N_y),A}\bra{m_1,m_2,B} +h.c.
        \end{aligned}
 \end{equation}

\begin{figure}[htpb]
    \begin{minipage}{0.55\textwidth}
         \includegraphics[width=0.36\linewidth,height=0.3\linewidth]{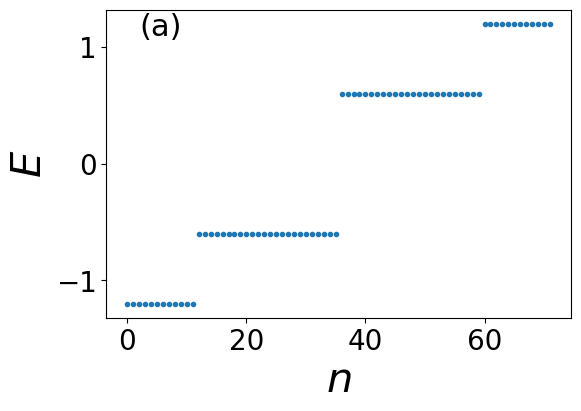} 
        \hspace{5pt}
        \includegraphics[width=0.36\linewidth,height=0.3\linewidth]{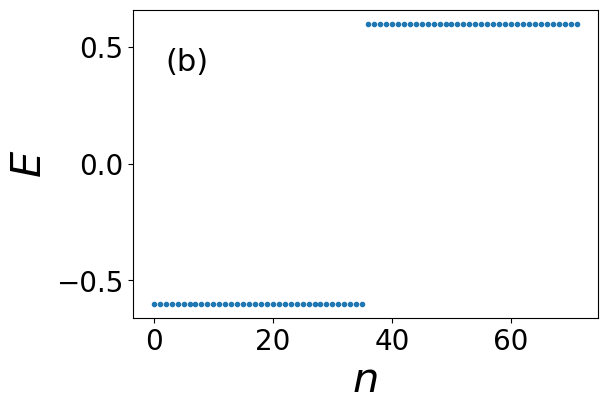}
  \end{minipage}
    \caption{The energy spectrum of the 2D extension topological atomic  model's Hamiltonian when the system is considered to be in CPBC in real space. The energy spectrum plotted on (a) is when the individual topological atomic  chains are in its one of the  extreme cases, where $v=0$ and  $w \neq 0$ with $t=0$. The energy spectrum plotted on (b) is when the individual topological atomic  chains are in its other extreme cases, where $w=0$ and $v \neq 0$ with  $t<\frac{v+w}{2}$. No zero modes  appear in either of the cases, confirming that all the zero modes appearing in Figure  \ref{fig:8} are emerging from the boundary.}
    \label{fig:9}
\end{figure}

The energy spectrum for the extreme cases depicts the absence of the zero modes, as shown in Figure \ref{fig:9}, conforming that they are solely the result of boundary effects. Moreover, the energy spectrum is symmetric about zero because of the CS and PHS. Summing up, for a 2D system we can have distinct boundary effects depending on the periodic condition we place on the system.  A PPBC may explain the impact of the boundary, but it is only complete if the analysis also considers the OBC to see the full range of boundary effects. The choice of bulk is irrelevant, but the choice of boundary plays a significant role in establishing a bulk-boundary correspondence in these kinds of unconventional stacking. The presence of numerous zero modes in its non-trivial topological phase, when the system is in OBC, is the evidence of identifying a crucial boundary to establish bulk-boundary correspondence, which was missing in the previous studies \cite{43.Master_thesis}.

\section{Bulk-Boundary Correspondence}
\label{sec: Bulk-Boundary Correspondace}

\begin{figure}[htbp]
    \centering
    \includegraphics[width=0.5\textwidth]{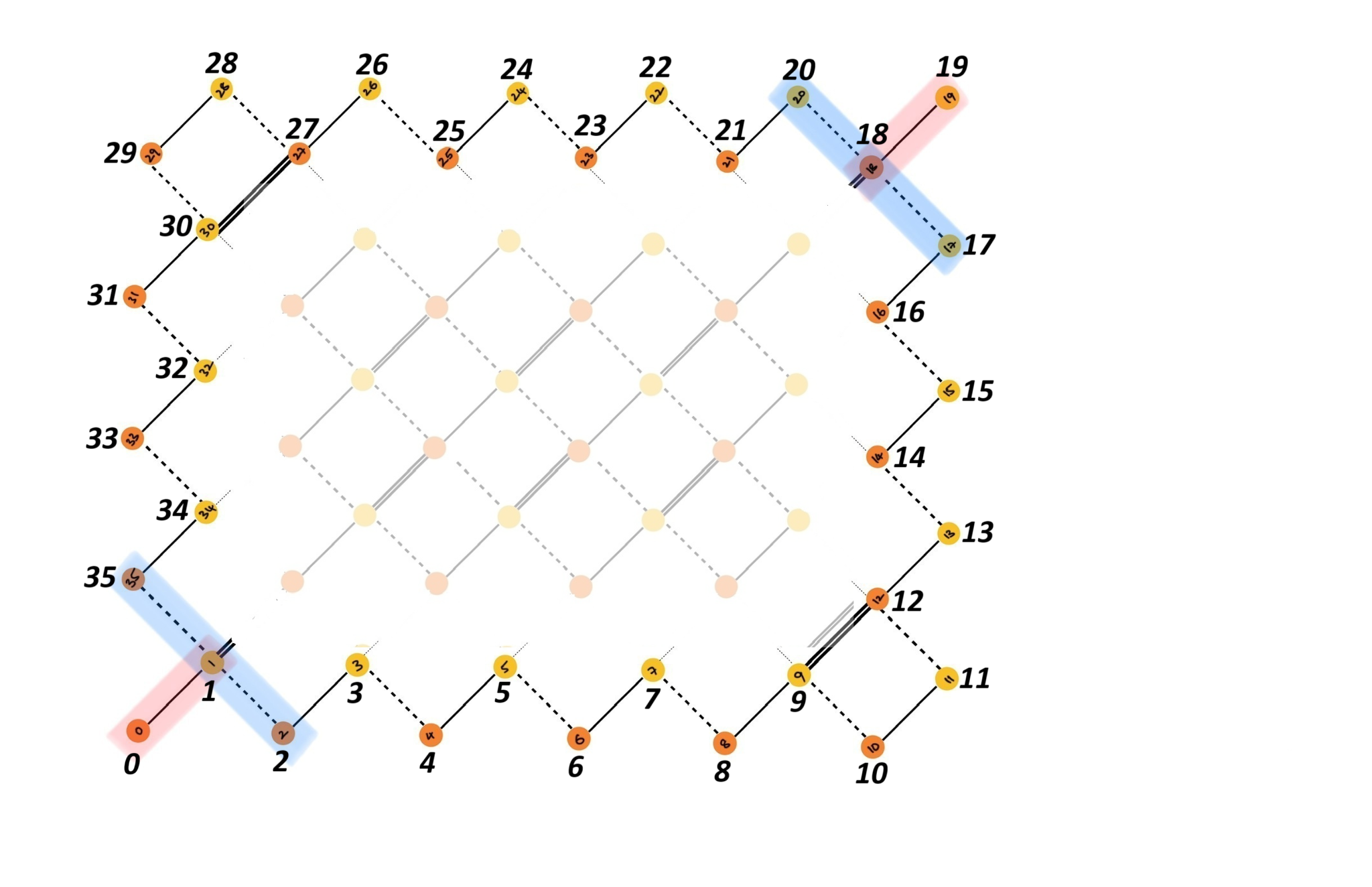}
    \caption{A pictorial illustration of the peripheral layer of the system with the total number of lattice points $18$  when $N_x=6$ and $N_y=3$ in the 2D model. Two effective topological atomic  chains share this $18$ lattice points. The intersection of these chains results in domain wall formation in terms of dimer (red) and trimer (blue). Atoms A and B are shown by orange and yellow dots, with intra-chain hopping $v$  by thin lines and inter-chain hopping $t$ denoted by dashed lines, respectively.}
    \label{fig:10}
\end{figure}

So far, the last two sections have thoroughly examined the system's bulk and attempted to understand its boundary effects. This section establishes a correspondence between the bulk topological invariant and the zero modes at the boundaries. Figure \ref{fig:10} depicts the peripheral layer of the system where two effective topological atomic  chains intersect, resulting in topological defects.

The peripheral layer has $18$ lattice points with $N_x=6$ and $N_y=3$ in the 2D model. These lattice points are shared by two atomic  chains with defects. These two chains are connected together at two points using two dimers, resulting in two domain walls composed of dimers and trimers. Each chain has two atoms, A and B, per lattice, with intra-cell hopping $v$ and inter-cell hopping $t$. 
Each chain also allows a NNN hopping at one site, potentially in the center of the peripheral chain if $N_x=N_y$, while maintaining the system's chiral symmetry. If $N_x>N_y$, the site advances to the right from the center of the chain; otherwise, it moves to the left.  The location of this site does not affect the domain walls.  The Hamiltonian of the peripheral chain is as follows:

\begin{equation}
    \hat{H}=\hat{H}_{\text{eff}}+\hat{H}_{\text{dimer}}+\hat{H}_{\text{eff-dimer}},
\end{equation}
where $\hat{H}_{\text{eff}}$=$\hat{H}^{(1)}_{\text{eff} }+\hat{H}^{(2)}_{\text{eff}}$,
$\hat{H}_{\text{dimer}}$= $\hat{H}^{(1)}_{\text{dimer}}+\hat{H}^{(2)}_{\text{dimer}}$, and finally, 
$\hat{H}_{\text{eff-dimer}}$ represents the coupling between $\hat{H}_{\text{eff}}$ and $\hat{H}_{\text{dimer}}$. Diagonalizing the above Hamiltonian gives the energy spectrum and the energy eigenfunction.

\begin{figure}[htbp]
    \begin{minipage}{0.5\textwidth}
         \includegraphics[width=0.45\linewidth,height=0.4\linewidth]{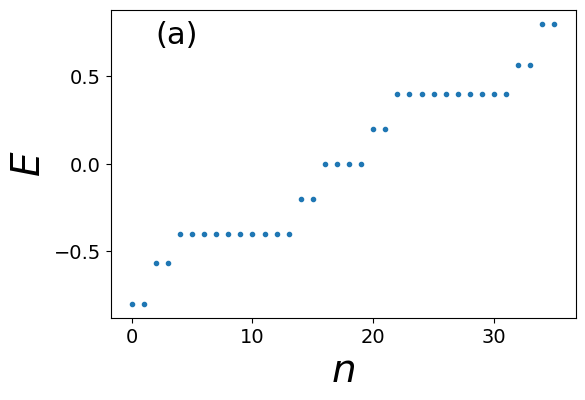} 
        \hspace{5pt}
        \includegraphics[width=0.45\linewidth,height=0.4\linewidth]{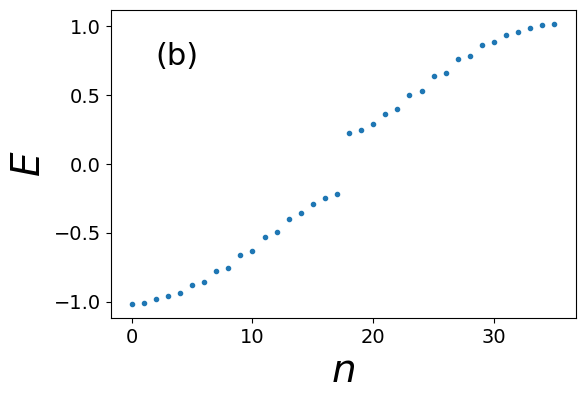}
  \end{minipage}
    \caption{The energy spectrum of the peripheral chain  of the 2D model (a) has four zero modes when the chain is in the non-trivial topological phase as $v=0$ and  $w \neq 0$, with $t$ to be a small perturbation and with $t<\frac{v+w}{2}$. The energy spectrum (b) has no zero modes when the boundary chain is in the trivial topological phase with $t<v$ and  $t<\frac{v+w}{2}$, $v\neq0$ and $w \neq0 $.}
    \label{fig:11}
\end{figure}

Figure \ref{fig:11} identifies four zero modes when the peripheral chain is in its non-trivial topological phase with $v=0$, $w$, and $t$ not equal to zero. The conditions imposed on $v,w$, and $t$ on the chain also satisfy the conditions of the 2D model in its non-trivial topological phase as $v=0$ and  $w \neq 0$, with $t$ to be a small perturbation and with $t<\frac{v+w}{2}$. No zero modes appeared when the boundary chain is in the trivial topological phase with $t<v$ and $t<\frac{v+w}{2}$, $v\neq0$ and $w \neq0 $, which are also the conditions from the 2D system to be in trivial topological phases.

 \begin{figure}[htpb]
    \begin{minipage}{0.55\textwidth}
  \includegraphics[width=0.35\linewidth,height=0.3\linewidth]{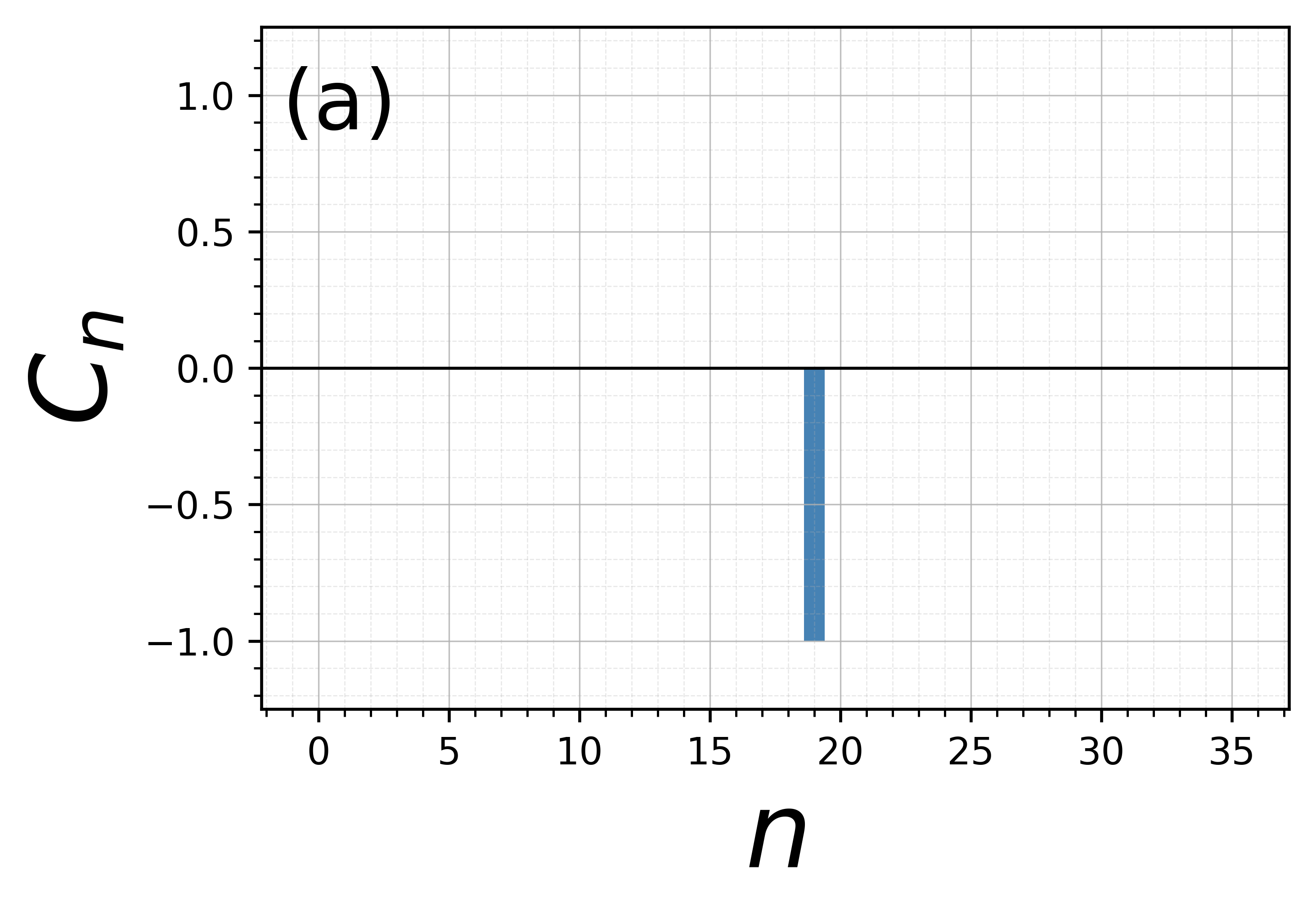} 
  \hspace{10pt}
  \includegraphics[width=0.35\linewidth,height=0.3\linewidth]{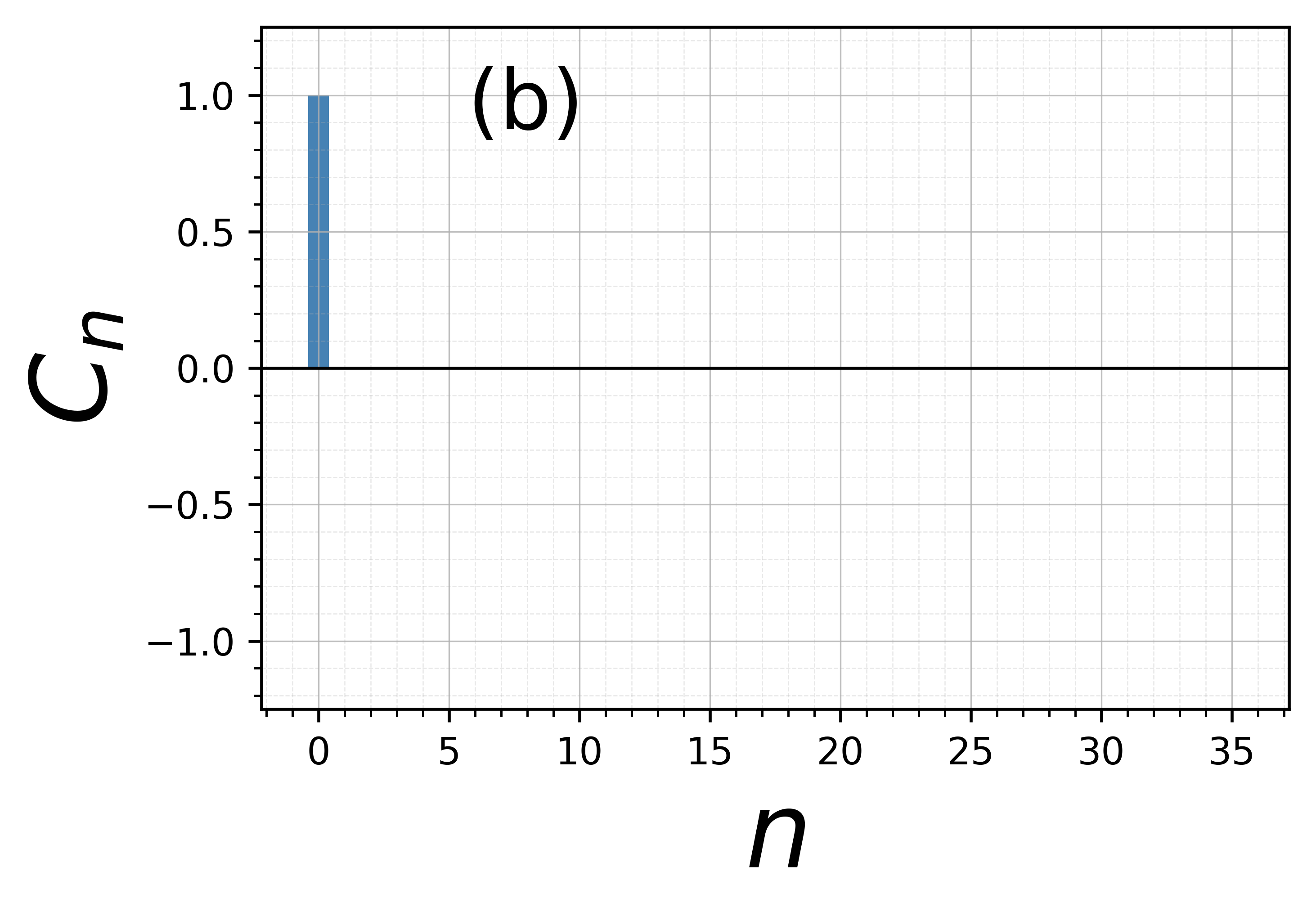}
\end{minipage}

\begin{minipage}{0.55\textwidth}
  \includegraphics[width=0.35\linewidth,height=0.3\linewidth]{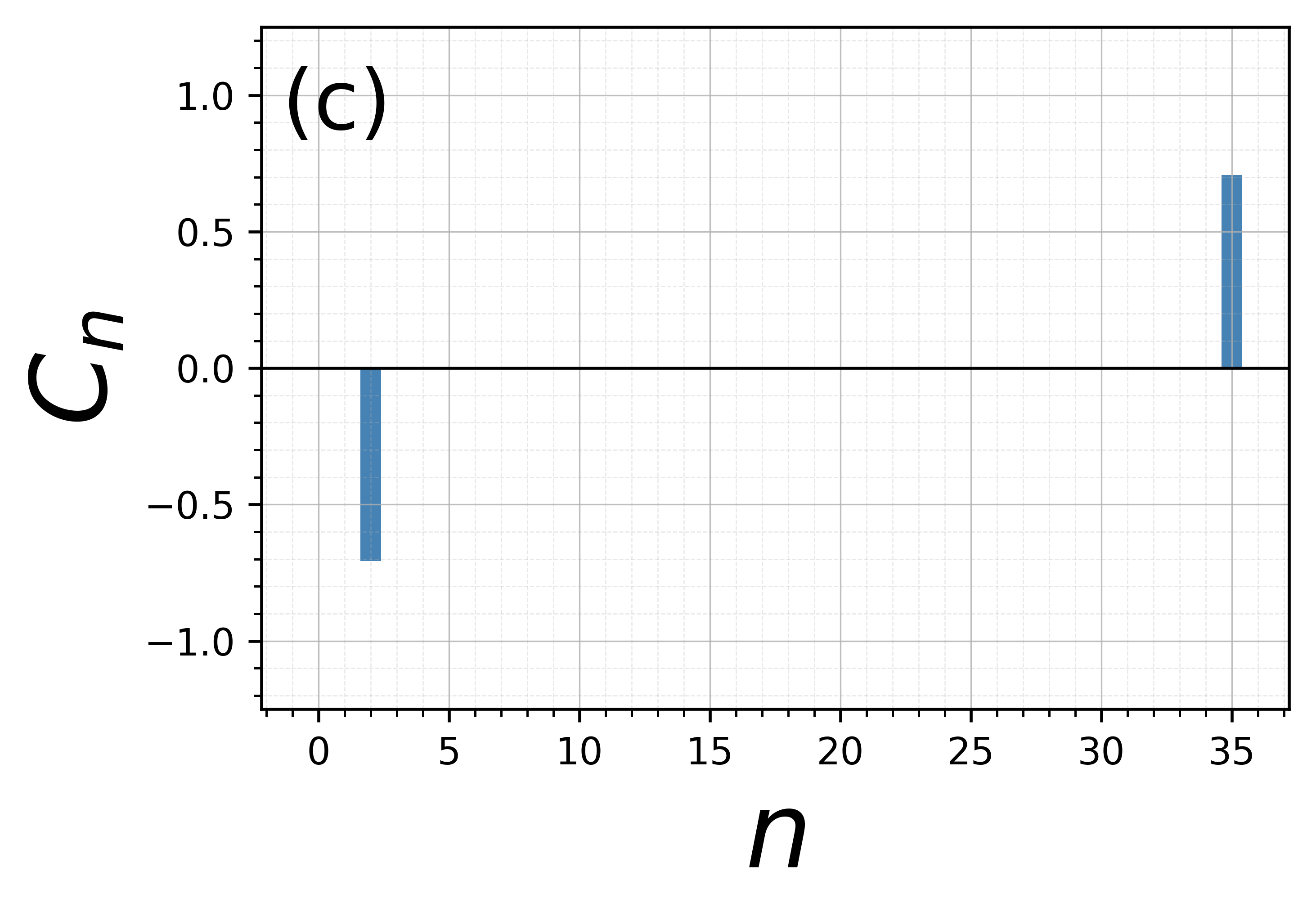} 
  \hspace{10pt}
  \includegraphics[width=0.35\linewidth,height=0.3\linewidth]{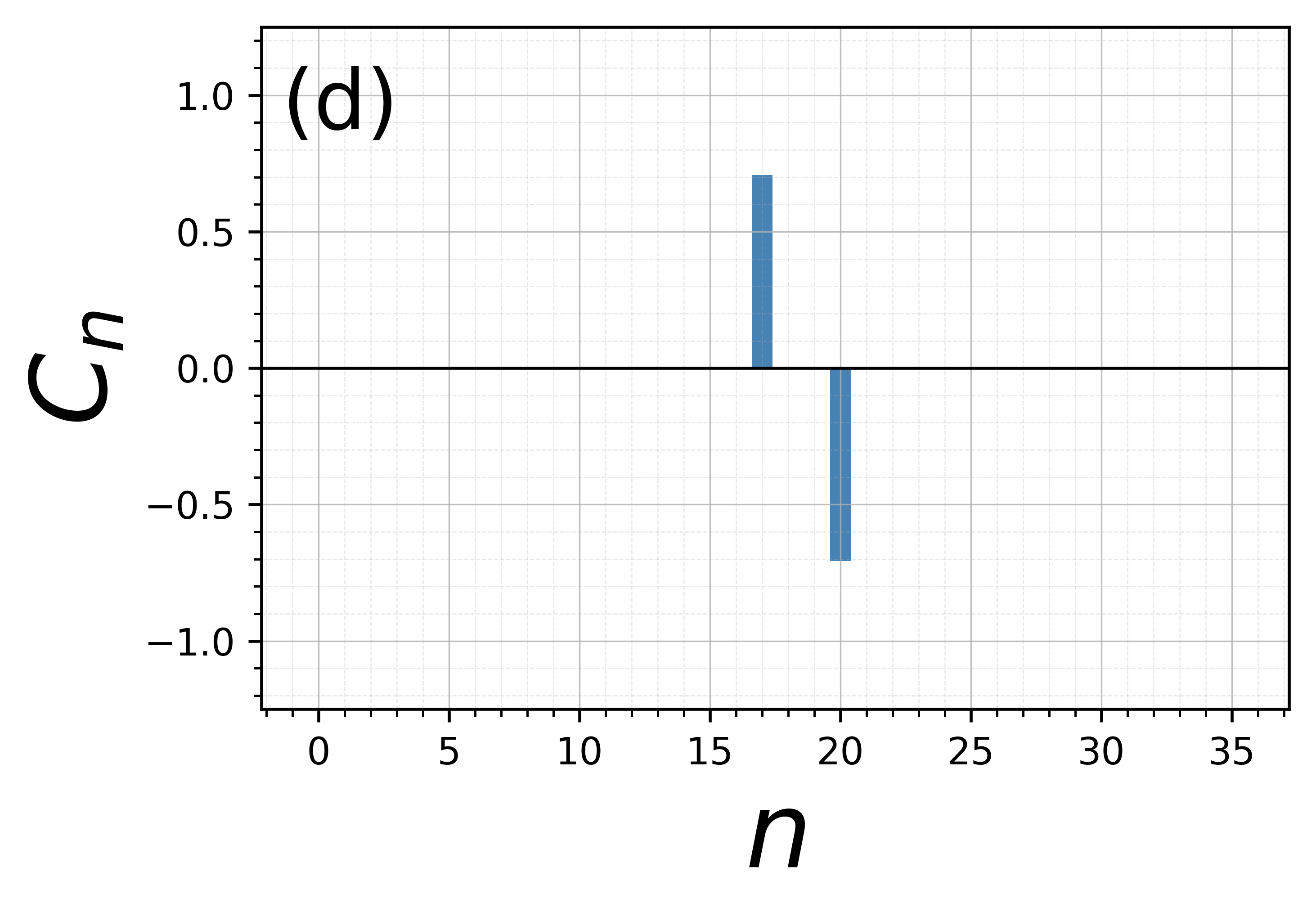}
\end{minipage}

    \caption{Schematic representation of the energy eigenfunctions of the four zero modes of the outermost chain is considered. These plots represent the amplitudes of the energy eigenfunctions ($C_{n}$) with spatial distribution when the 2D model is in a non-trivial topological phase as $v=0$ and  $w \neq 0$ with $t$ to be a small perturbation and $t<\frac{v+w}{2}$. Plots (a) and (b) indicate the amplitudes of energy eigenfunctions contributing from the locations 19(a) and 0(b), which are the isolated sites B and A, respectively. Plots (c) and (d) represent the amplitudes of energy eigenfunctions contributing from two specific locations at 2 and 35 (c) and 17 and 20(d), respectively. It indicates that two zero modes are associated with two trimers.}  
    \label{fig:12}
\end{figure}
\begin{figure}[htpb]
   \begin{minipage}{0.55\textwidth}
       \includegraphics[width=0.36\linewidth,height=0.3\linewidth]{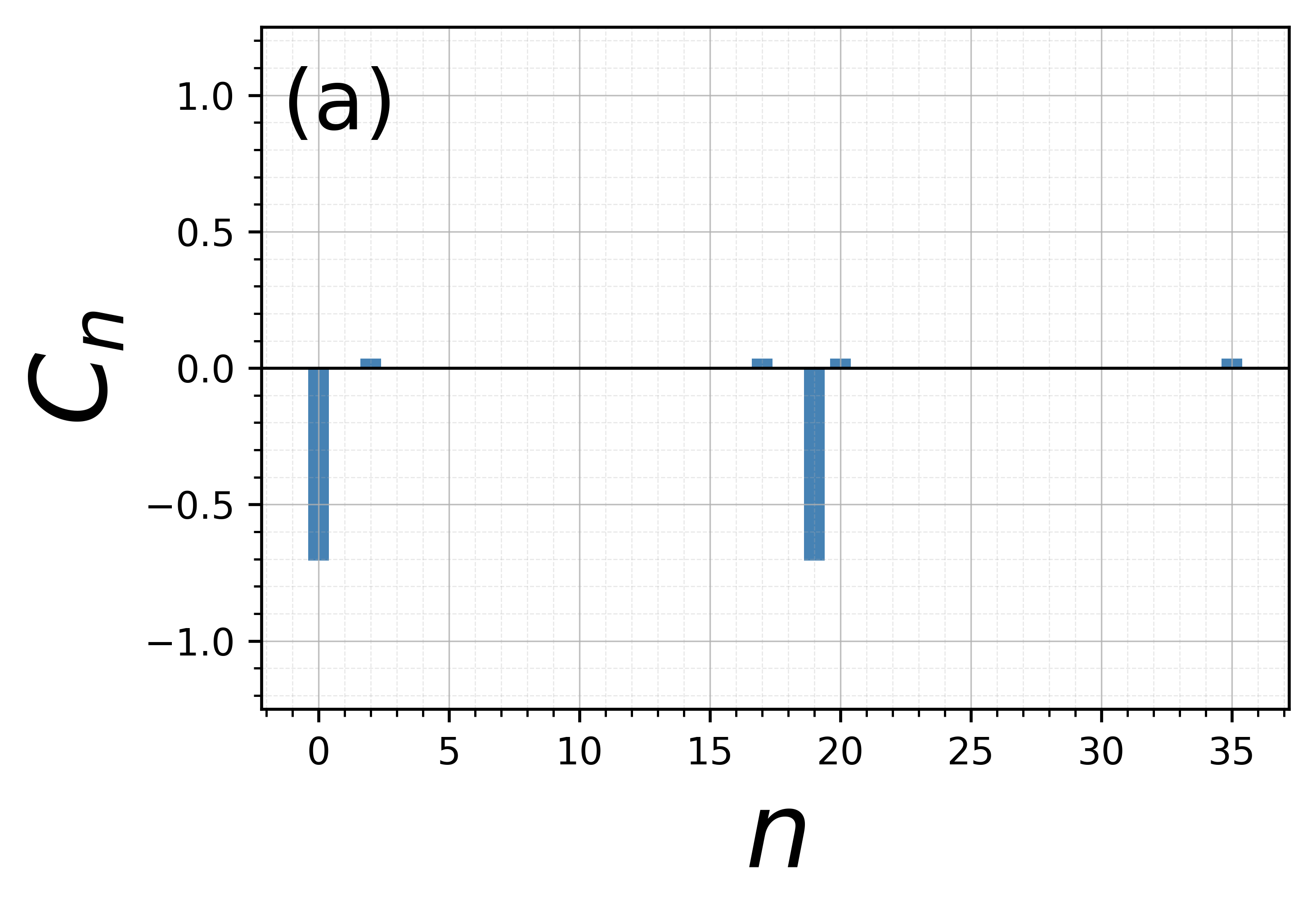}
       \includegraphics[width=0.36\linewidth,height=0.3\linewidth]{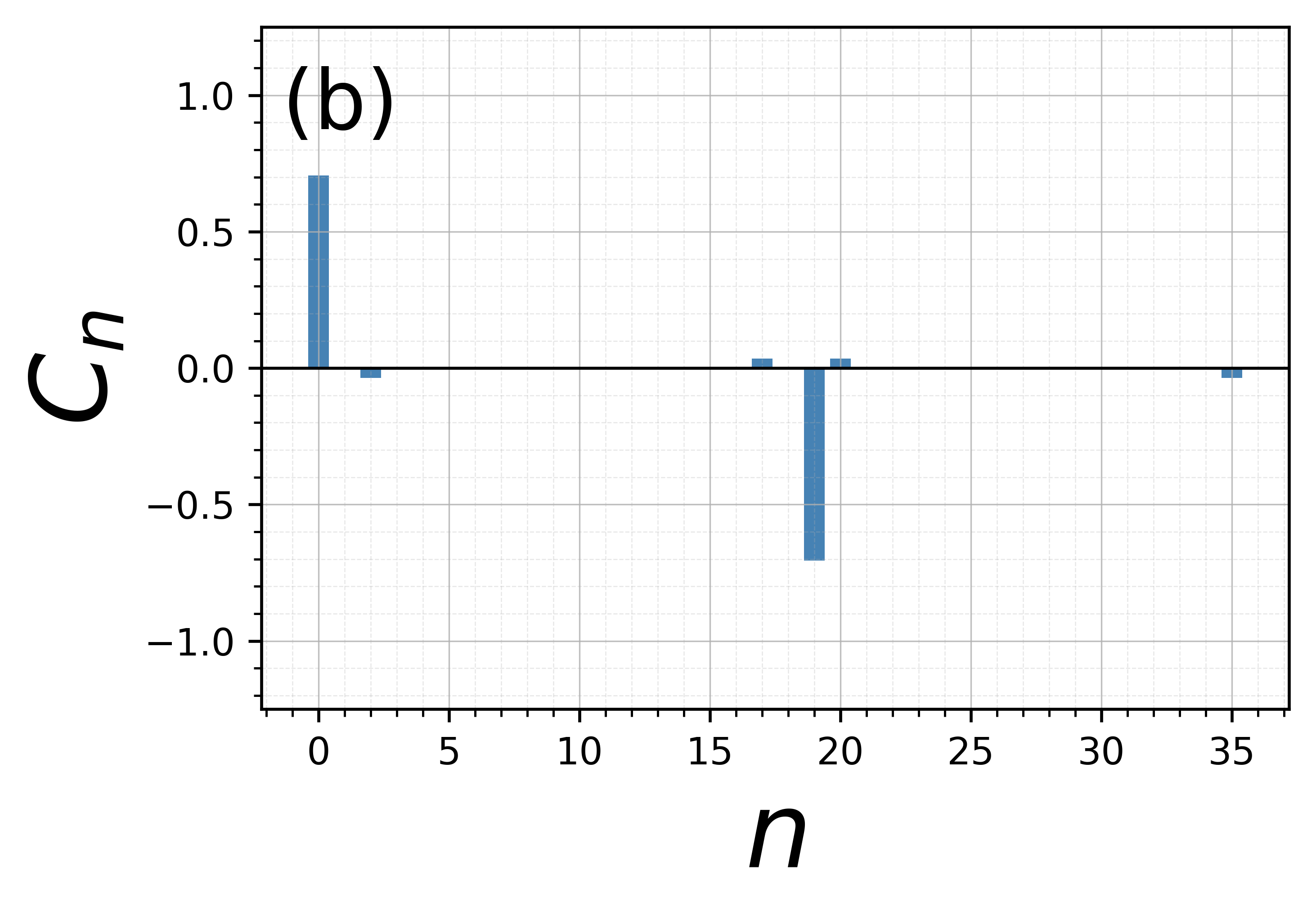}
   \end{minipage}
    \caption{Schematic representations of the amplitude energy eigenfunctions ($C_{n}$) of the two lowest energy eigenvalues (of order $10^{-11}$ in the units of $t$) when the outermost chain of the boundary is considered. These plots represent the amplitudes of the energy eigenfunctions with spatial distribution when 2D system satisfies with $t>\frac{v+w}{2}$,  $v$  and $w$ less than $t$.  Both (a) and (b) have the amplitudes from the site locations $0$ and $19$ which represent the sites A and B.} 
    \label{fig:13}
\end{figure}

Figure \ref{fig:12} shows the spatial distribution of the energy eigenfunctions for the four zero modes. These four zero-mode eigenfunctions correspond to two trimers and two isolated atoms in the peripheral chain. The two eigenfunctions localize on the A and B sites, with spatial coordinates $0$ and $19$. If the trimer is formed by $\boldsymbol{(A-B-A)}$ bonding with spatial locations $2$ and $35$, the eigenfunction only contributes from site A. If the trimer is formed by $\boldsymbol{(B-A-B)}$ bonding, the eigenfunction contributes exclusively from site B, which has spatial positions $17$ and $20$, as seen in Figures \ref{fig:10} and \ref{fig:12}. Change in  polarization along $P_{x}$ and $P_{y}$, as well as crystalline mirror symmetry along the $y=x$ line, creates a confined charge shift along the $y=x$ corners. Because of this effect, the amplitude $C_{n}$ shown in Figure \ref{fig:12} of (c) and (d) is equally distributed on the  A  and B sites for $\boldsymbol{(A-B-A)}$ and  $\boldsymbol{(B-A-B)}$ bonding, respectively, resulting in trimers formation.  As a consequence, the zero modes naturally cannot be restrained on one side, but must appear symmetrically around the mirror-symmetric line. Further localization on either $\boldsymbol{A}$ or $\boldsymbol{B}$ sub-lattices in the emergent trimer structures $\boldsymbol{(A-B-A)}$ and $\boldsymbol{(B-A-B)}$ is consistent with the constraints imposed by CS, which prohibits localization between sub-lattices. This demonstrates how crystalline mirror-symmetry protects the topological states and also directs their spatial distributions, providing a design principle for creating symmetry-guided localized synthetic quantum materials. When $t>\frac{v+w}{2}$ with $v<w$ and $v>w$ conditions are applied to the peripheral chain, the lowest eigenvalue is in the order of $10^{-11}$ when $t=1$ with $v=0.2$ and $w=0.1$ or $v=0.1$ and $w=0.2$. The related eigenfunctions are localized to isolated sites A and B, as shown in Figure \ref{fig:13}.\\

When the system is in a non-trivial topological phase, where $v<w; t<\frac{v+w}{2}$, with a bulk topological invariant $\nu=2$, the four zero-energy modes on the trimers $\boldsymbol{(A-B-A)}$,  $\boldsymbol{(B-A-B)}$ and the isolated atoms, it aids in establishing the bulk-boundary correspondence in the system. The other non-trivial topological phases, where $v<w; t>\frac{v+w}{2}$ and $v>w; t>\frac{v+w}{2}$ with $\nu=1$, support two modes of order $10^{-11}$, which are associated with the isolated sites A and B, also establish the bulk-boundary correspondence.

\begin{center}
\begin{table}[htpb]
    \begin{tabular}{ ||p{2.3cm}|p{1.2cm}|p{1cm}|p{1cm}|p{0.6cm}|| }
 \hline
 {} & \multicolumn{3}{c|}{Zero modes}&{} \\
 \hline 
 Topological phases & Trimers & Isolated atoms  & Total  & $\nu$ \\
 \hline \hline
$v<w; t<\frac{v+w}{2}$ & 2    & 2 & 4 & 2 \\
\hline
 $v<w; t>\frac{v+w}{2}$ & 0    & 2 & 2 & 1 \\
 \hline
 $v>w; t>\frac{v+w}{2}$ &  0    & 2 & 2 & 1 \\
 \hline
 $v>w; t<\frac{v+w}{2}$ & 0    & 0 & 0 & 0 \\
 \hline
 
\end{tabular}
    \caption{A summary of the topological phase categorization in the 2D system. The table's extreme left column shows the specific conditions on $v$, $w$, and $t$ for the appearance of various topological phases. The table shows the bulk topological invariant winding number ($\nu$) in the far right column. The column zero modes show the overall number of zero modes in the system and the individual contributions of the trimers and isolated atoms for the zero modes in the system, respectively.}
    \label{tab:1}
\end{table}
\end{center}

\section{Conclusion}
\label{sec:conclusion}
In conclusion, we look for a general way to identify a suitable lower-dimensional topological invariant (winding number) to classify a higher-order topological material. The model considered here stacks the topological atomic  chains in two dimensions with topological defects at the corners. Here, we study the phase evolution of the Bloch states and identify a set of closed curves in the parametric space $(k_x,k_y)$, which belong to the same homotopy group. These curves will produce symmetry-protected closed curves in the vector space of $(h_x,h_y)$, which help us distinguish the system's topological phases in terms of non-zero winding numbers. This study aligns with previous attempts \cite{26.Khalaf_2018,73.PhysRevLett.123.177001} on extending lower-dimensional topological systems into higher dimensions to explore higher-order topological phases. Signatures of a typical higher-order topological system are the occurrence of lower-dimensional localized zero modes.
Interestingly, in this system, because of the non-trivial nature of stacking as elaborated in Section \ref{sec: The 2D extension of the topological atomic  model}, we find the appearance of zero energy-modes at two of the corners in the form of topological defects (trimers) and isolated atoms formed at the intersection of a pair of effective SSH chains at the boundaries. These zero-energy states  are protected by crystalline mirror-symmetry and chiral symmetry. Association of these zero modes with the non-zero bulk winding numbers and the diagonal polarization results in interesting bulk-boundary (corner) correspondence in the second-order topological phase of this system.\\

 Recent experimental observation supports our finding of corner modes that are substantially localized at defective edges, as shown in this work \cite{Cheng2025}. However, experimental realization of the topological atomic  model in electric circuits, photonics, engineered optical wave guides \cite{74.PhysRevResearch.3.023056,75.PhysRevResearch.4.013185,76.Longhi:18,77.Longhi:13} in realization of higher-order phases  \cite{78.SerraGarcia2018,79.PhysRevLett.122.233903,80.PhysRevLett.122.233902,81.PhysRevB.99.020304,82.Mittal2019} have been successfully developed in recent times. In principle, such studies can help us to have an experimental realization of the model we study here; it will be fascinating to explore further, both theoretically and experimentally, the correspondence between the lower-dimensional topological invariant and topologically protected defects states (trimers) at the boundaries in terms of quantum entanglements \cite{Wen2002,Bahri2015,PhysRevB.82.241102}.\\\\

\section*{acknowledgment}
Authors gratefully acknowledge Saurabh Basu, (IIT Guwahati) for his thorough and insightful revision of manuscript.\\

Authors also acknowledge the valuable discussions with  A. P. Balachandran (Syracuse University), Rakesh Tibrewala (The LNMIIT), and Pankaj Bhalla (SRM University-AP). A special thanks to Jasrah Farooq (The LNMIIT) for copy editing the manuscript and Abhishek Joshi (Brandenburg Tech. University) for his association with the authors at the initial stages of the work.

\bibliography{2d-ssh-Ref}

\end{document}